\documentclass[12pt]{article}
\usepackage{amsmath,amssymb,amsthm,amssymb,amsfonts,mathtools,bm,algorithmic,algorithm,stmaryrd}
\usepackage{booktabs,latexsym,CJK,multirow,multicol,caption,enumerate,geometry,float,appendix,endnotes,setspace,tabularx}
\usepackage{graphicx,color,times,hyperref,natbib}
\usepackage[english]{babel}
\usepackage{subcaption}
\usepackage[T1]{fontenc}
\newtheorem{thm}{Theorem}

\newtheorem{prop}{Proposition}
\newtheorem{corollary}{Corollary}

\newtheorem{example}{Example}

\def\mb{\mathbb}
\def\mf{\mathbf}
\def\mc{\mathcal}

\def\tc{\tilde{c}}

\def\bb{\bm{b}}
\def\bd{\bm{d}}
\def\be{\bm{e}}

\def\bB{\bm{B}}
\def\bC{\bm{C}}
\def\bD{\bm{D}}
\def\bG{\bm{G}}
\def\bH{\bm{H}}
\def\bJ{\bm{J}}

\def\bV{\bm{V}}
\def\bW{\bm{W}}
\def\bX{\bm{X}}

\def\bbeta{\bm{\beta}}

\def\bmu{\bm{\mu}}
\def\bnu{\bm{\nu}}
\def\bomega{\bm{\omega}}
\def\bpsi{\bm{\psi}}
\def\bvartheta{\bm{\vartheta}}
\def\bth{\bm{\theta}}
\def\bzeta{\bm{\zeta}}
\def\bGamma{\bm{\Gamma}}
\def\bLambda{\bm{\Lambda}}

\def\bPsi{\bm{\Psi}}
\def\bTh{\bm{\Theta}}
\def\mE{\mb{E}}
\def\mR{\mb{R}}
\def\cK{\mathcal{K}}
\def\cL{\mathcal{L}}

\def\bcA{\bm{\mathcal{A}}}
\def\bcB{\bm{\mathcal{B}}}

\def\bcW{\bm{\mathcal{W}}}
\def\bYt{\bm{Y}(t)}
\def\bXt{\bm{X}(t)}
\def\P{\textrm{Pr}}
\def\vec{\textrm{vec}}
\def\trans{^{\mbox{\scriptsize {\sf T}}}}
\def\cp{\llbracket \bm{\nu}; \bm{B}^{y}, \bm{B}^{x},\bm{B}^{c} \rrbracket}
\newcommand{\argmin}{\arg\!\min}

\begin{document}

\title{\Large{\textbf{Multivariate Temporal Point Process Regression}}}
\medskip

\author{\large{Xiwei Tang and Lexin Li}}
\date{}
\maketitle

\begin{footnotetext}[1]
{Xiwei Tang is Assistant Professor, Department of Statistics, University of Virginia. Email: xt4yj@virginia.edu. 
Lexin Li is Professor, and the corresponding author, Department of Biostatistics and Epidemiology, University of California, Berkeley. Email: lexinli@berkeley.edu}
\end{footnotetext}

\baselineskip=20pt

\begin{abstract}
Point process modeling is gaining increasing attention, as point process type data are emerging in numerous scientific applications. In this article, motivated by a neuronal spike trains study, we propose a novel point process regression model, where both the response and the predictor can be a high-dimensional point process. We model the predictor effects through the conditional intensities using a set of basis transferring functions in a convolutional fashion. We organize the corresponding transferring coefficients in the form of a three-way tensor, then impose the low-rank, sparsity, and subgroup structures on this coefficient tensor. These structures help reduce the dimensionality, integrate information across different individual processes, and facilitate the interpretation. We develop a highly scalable optimization algorithm for parameter estimation. We derive the large sample error bound for the recovered coefficient tensor, and establish the subgroup identification consistency, while allowing the dimension of the multivariate point process to diverge. We demonstrate the efficacy of our method through both simulations and a cross-area neuronal spike trains analysis in a sensory cortex study.
\end{abstract}


\newpage

\section{Introduction}
\label{sec:introduction}

Point process is drawing increasing attention, as data in the form of point process are emerging in a wide variety of scientific and business applications. Examples include forest ecology \citep{stoyan2000}, spatial epidemiology \citep{diggle2010risk}, social network modeling \citep{perry2013}, neuronal activity modeling \citep{brown2004, chen2019}, functional neuroimaging meta analysis \citep{Kang2011, Kang2014}, among others. In general, a  point process is a collection of events, or points, randomly located in some domain space, e.g., a spatial domain or a time domain. Our motivation is a neuronal spike trains analysis in a sensory cortex study \citep{okun2015}. A newly developed two-photon calcium imaging technique is now greatly facilitating neuroscience studies, by enabling  simultaneously recording of the dynamic activities for a population of neurons while maintaining individual neuron resolution \citep{ji2016}. In our study, there are 139 and 283 neurons imaged simultaneously from two areas of a rat's brain, the primary visual cortex area (V1) and the primary auditory cortex area (A1). In a visual activity, it is known that some locations of the primary visual cortex would respond to input from auditory and other sensory areas \citep{liang2013}. One of the scientific goals of this study is to understand the association patterns and information transmissions between the neurons across A1 and V1, and to model potential excitation or inhibitory effects of neuron firings between the two areas. 

In this article, we propose a new multivariate point process regression model to address this question, where both the response and the predictor can be a high-dimensional point process. We model the predictor effects through the conditional intensities using a set of basis transferring functions in a convolutional fashion.  We organize the corresponding transferring coefficients in the form of a three-way tensor,  then impose the low-rank, sparsity, and subgroup structures. Both low-rank and sparsity are commonly used low-dimensional structures in high-dimensional data analysis, and are scientifically plausible in neuroscience and many other applications \citep{zhou2013tensor, ChenYuan2019, ZhangA2019, Bacry2020}. Subgroup is another frequently used structure in plenty of applications, and it corresponds to ensemble neural activities in neuroscience \citep{okun2015}. Together these structures effectively reduce the number of free parameters, and greatly facilitate the model interpretation. Under the proposed model, we develop a highly scalable alternating direction method of multipliers algorithm for parameter estimation. We establish the asymptotic properties of the penalized maximum likelihood estimator while allowing the dimensions of both the response and the predictor process to diverge. We also comment that, although motivated by a neuroscience problem, our method is equally applicable to numerous other point process applications, e.g., the social infection network learning \citep{zhou2013network}. 

There have been a large number of models targeting a spatial point process, where its local intensity is usually assumed to depend on some deterministic features or some location-relevant random variables \citep[see, e.g.,][]{guan2008consistent, guan2015quasi, Kang2014, GuanZhang2017}. We instead aim at a temporal point process, which is usually evolutionary in nature, in that the occurrence of a future event depends on the historical realizations of the process. This leads to a different set of model assumptions and modeling techniques. Moreover, most classical inhomogeneous point process solutions target a univariate or bivariate process along with a limited number of predictors \citep{diggle2010risk, waagepetersen2009two}. We instead target high-dimensional response and predictor processes, and we allow the dimension of both processes to diverge. This has introduced new challenges in both modeling and theoretical analysis. 

There have also been a family of models targeting a temporal point process, all of which are built on a self-exciting process called the Hawkes process \citep{hawkes1971}. A Hawkes process assumes that a future event is triggered by its own past events, and is widely used in neuronal spike trains analysis as well. In recent years, there have been a number of point process models extending the Hawkes process. Our proposal is related to but also clearly distinctive from the existing models in multiple ways. First of all, our model extends the classical Hawkes process, by simultaneously incorporating nonlinear and inhomogeneous intensities, multiple basis functions, diverging point process dimensions, and additional structures on the transferring coefficients. On the other hand, our model is more general, in that it allows a wide class of stochastic processes to be predictors, and can be applied to the scenarios where the Hawkes process is applicable, but not vice versa. We give some examples in Section \ref{sec:regmodel}. Second, \cite{zhou2013network} introduced low-rank and sparsity structures in a multivariate Hawkes model.  \cite{Bacry2020} studied the theoretical properties of the model, while potentially allowing the point process dimension to diverge. Our proposal employs similar 
low-dimensional structures and explicitly studies the diverging dimension. However, there are numerous fundamental differences. \cite{zhou2013network, Bacry2020} both imposed the linear link function and the stationary assumption, whereas we consider a general and potentially nonlinear link, and do not require the process to be stationary. In addition, \cite{zhou2013network, Bacry2020} organized the transferring coefficients in a matrix form and placed a low-rank structure on the coefficient matrix, whereas we organize the transferring coefficients in a tensor form and employ a low-rank tensor decomposition. Tensor decomposition is considerably different from matrix decomposition \citep{kolda2009}, and naively transforming a tensor to a matrix may lose information. More importantly, the estimation error bound obtained by \cite{Bacry2020} is to increase as the point process dimension diverges, and as such the estimator is to suffer from the increasing dimension. By contrast, in Section \ref{sec:convergence}, we show that the diverging dimension is to benefit our penalized maximum likelihood estimator, and leads to a faster convergence rate on the asymptotic concentration. Third, \cite{hansen2015} considered an intensity-based model for multivariate Hawkes process with a fixed dimension, and adopted the least-squares estimation and $\ell_1$ regularization. However, the key difference is that \cite{hansen2015} modeled each individual point process separately, while we model multiple processes jointly.  More specifically, even though \cite{hansen2015} targeted a multivariate point process, their loss function was completely separable. As such, they essentially modeled each individual point process one at a time. By contrast, we model multiple response point processes in a joint fashion, in that we integrate and borrow information across different response processes. This is achieved through both the tensor latent factors that are shared across all intensity functions, as well as the subgrouping structure on the transferring coefficients that encourage information sharing among similar individual processes. Such a joint modeling strategy essentially leads to the improved estimation bound when the number of response processes increases. Finally, \cite{bacry2016} and \cite{chen2019} studied some multivariate Hawkes process models using moment-based statistics, while we focus on modeling the conditional intensity function. In Section \ref{sec:regmodel}, we discuss in more detail why the intensity-based approach is more suitable than the moment-based approach in our setting.

The rest of the article is organized as follows. Section \ref{sec:model} introduces our proposed multivariate temporal point process regression model. Section \ref{sec:estimation} develops the estimation algorithm, and Section \ref{sec:theory} derives the theoretical properties. Section \ref{sec:simulations} presents the simulations, and Section \ref{sec:realdata} illustrates with a neuronal spike trains data analysis. All proofs are relegated to the Supplementary Appendix.

\section{Model}
\label{sec:model}

\subsection{Background}
\label{sec:background}

We begin with a brief review of temporal point process, and we refer to \cite{daley2007} for more details. Specifically, a temporal point process is a stochastic counting process defined on the positive half of the real line $\mR^+$, and taking non-negative integer values. For a univariate process $X(t)$, let $t_1, t_2, \ldots \in \mR^+$ denote the event times, under which $X(A)=\sum_{l=1} \mf{1}_{[t_l \in A]}$ for any $A \in \mc{B}(\mR^+)$, and $\mc{B}(\mR^+)$  denotes the Borel $\sigma$-algebra of $\mR^+$. Define its mean intensity function as $\Lambda(t) = \lim_{dt \rightarrow 0}\mE[dX(t)]/dt$, where $dX(t)=X\big([t,t+dt)\big)$, and $dt$ is an arbitrary small increment of time.  A temporal point process is homogeneous if its mean intensity is a constant, and is inhomogeneous otherwise. If $\Lambda(t)$ is also a stochastic process, then it is a doubly stochastic process; e.g., a Cox process.  A temporal point process is usually assumed to be orderly; i.e., $\P(dX(t) > 1)=o(dt)$, which implies that $\Lambda(t) = \P(dX(t) =1)$. In addition, a point process $X(t)$ is stationary, if the distribution of $dX(t)$ only depends on the length of $dt$ but not the location $t$ on the time line. It is straightforward to generalize the notion of a univariate point process to a multivariate point process, i.e., $\bX(t) = \big(X_{1}(t), \ldots,X_{p}(t)\big)\trans$.

Moment statistics are widely used in point process modeling, especially for a stationary process \citep{guan2008consistent, guan2011second, chen2019}. Considering a $p$-dimensional stationary point process $\bXt$, define its first-order moment statistic, i.e., the mean intensity, and its second-order statistic, i.e., the  covariance, as,
\begin{equation*} 
\begin{split}
\Lambda^x_i & =  \mE\{d X_i(t)\}/dt, \quad i=1, \ldots, p, \\
V^x_{ij}(\tau) & =  \mE\{d X_i(t)d X_j(t-\tau)\} / \{dt d(t-\tau)\}-\Lambda_i\Lambda_j-\delta_{ij}(\tau)\Lambda_i, \quad i, j = 1, \ldots, p,
\end{split}
\end{equation*}
respectively, where $\delta_{ij}(\tau)=0$ if $i \ne j$, and $\delta_{ij}(\tau)=\delta(\tau)$ if $i = j$, and $\delta(\cdot)$ denotes the Dirac delta function satisfying that $\delta(x)=0$ for $x \ne 0$ and $\int_{-\infty}^{+\infty} \delta(x)dx=1$. Write $\bLambda^x = (\Lambda^x_1, \ldots, \Lambda^x_p)\trans \in \mR^p$, and $\bV^{xx}(\cdot) = \big(V^x_{ij}(\cdot)\big): \mR \mapsto \mR^{p \times p}$.  Analogously, consider another $m$-dimensional stationary point process $\bYt$, with the mean intensity $\bLambda^y = (\Lambda^y_1, \ldots, \Lambda^y_m)\trans \in \mR^m$, and the covariance $\bV^{yy}(\cdot) = \big(V^y_{ij}(\cdot)\big): \mR \mapsto \mR^{m \times m}$.   The cross-covariance between $X_j(t)$ and $Y_i(t)$ is defined as, 
\begin{eqnarray*}
C^{xy}_{ji}(\tau)=\mE\{ dX_j(t)dY_i(t-\tau) \} / \{dtd(t-\tau)\}-\Lambda^x_j\Lambda^y_i, \quad i=1,\ldots,m, j=1,\ldots,p.
\end{eqnarray*}
Write $\bC^{xy}(\cdot) = \big(C^{xy}_{ji}(\cdot)\big) : \mR \mapsto \mR^{p \times m}$, and $\bC^{yx}(\cdot) = \big(C^{yx}_{ij}(\cdot)\big) : \mR \mapsto \mR^{m\times p}$. 

Conditional intensity function is another extensively used tool for modeling both spatial and temporal point process with additional covariates. For our proposed multivariate point process regression, we mainly target the conditional intensity function, as we detail in the next section.

\subsection{Multivariate point process regression}
\label{sec:regmodel}

We consider a temporal regression model with a $p$-dimensional predictor process $\bXt$ and an $m$-dimensional response process $\bYt$. Letting $\mc{H}_t$ denote the $\sigma$-algebra generated by $\{\bXt, \bYt\}$, then the $\mc{H}_t$-predictable intensity function $\lambda^{y}_i(t)$ of the $i$th response process $Y_i(t)$ is defined as,
\begin{eqnarray*} 
\lambda^y_i(t)dt = \P\left\{ dY_i(t)=1|\mathcal{H}_t \right\}, \quad i = 1, \ldots, m. 
\end{eqnarray*}
We assume this conditional intensity function takes the form,
\begin{equation} \label{eq:int_f}
\lambda^y_i(t)=\phi \left\{ \mu_i+ \sum_{j=1}^p \left( \omega_{ij} * dX_{j} \right) (t) \right\}, \quad i=1,\ldots,m,
\end{equation}
where $\phi(\cdot)$ is a link function that is possibly nonlinear, e.g., a rectifier function $\phi(x)=\max(0,x)$, or a sigmoid function $\phi(x)=e^{x}/(1+e^x)$,  $\mu_i$ is the background intensity, and $\omega_{ij}(\cdot): \mR^+ \mapsto \mR$ is the transferring function, $i=1,\ldots,m, j=1,\ldots,p$. Write $\bmu=(\mu_1, \ldots, \mu_m)\trans \in \mR^{m}$, and $\bomega = \big( \omega_{ij}(\cdot) \big) \in \mR^{m \times p}$.  
To account for potential evolution over time, we further assume that the transferring function $\omega_{ij}(\cdot)$ models the historical information of the predictor process $\bXt$ in a convolutional fashion, in that,
\begin{eqnarray} \label{eq:convolution}
\left( \omega_{ij}* dX_{j} \right) (t) = \int_{0}^{t}\omega_{ij}(\Delta) dX_{j} (t-\Delta). 
\end{eqnarray}
Similar formulation such as \eqref{eq:convolution} has been commonly used in the temporal point process literature \citep[e.g.,][]{hawkes1971, zhou2013network}. We also  note that, \eqref{eq:int_f} and \eqref{eq:convolution} together cover a fairly general class of models. We do no require a linear link function, nor the stationary condition. We  do not enforce $\omega_{ij}(\cdot)$ to be non-negative, as typically in the classical Hawkes process model, and therefore allow both ``exciting'' and ``inhibiting'' effects among different processes. In addition, we allow the predictor process $\bXt$ to take a general form. The convolution in \eqref{eq:convolution} actually works for both a stochastic predictor process and a deterministic predictor process, corresponding to a stochastic integral or a Stieltjes integral, respectively.  

We next outline a number of examples covered under our proposed model framework. 

\begin{example}\label{ex1}
Let $X_{j}(t)$, $j = 1, \ldots, p$, be a deterministic function on $[0,\infty)$, and $\bYt$ turns to be a multivariate inhomogeneous Poisson process with the deterministic mean intensity function $\lambda_i^y(t)$, $i = 1, \ldots, p$.  In a neuronal activity study, $\bXt$ may represent the designed stimulus signal. 
\end{example}

\begin{example}\label{ex2}
Let $X_{j}(t)$, $j = 1, \ldots, p$, be a stochastic point process, and $\bYt$ turns to be a multivariate Cox process or a doubly-stochastic process. If the transferring function $\omega_{ij}(x)$ takes the form such as a Dirac delta function $\delta(x-t)$, then the conditional intensity $\lambda_i^y(t)$ only depends on the value of $\bXt$ at  point $t$.  Consequently, our model can also be applied to non-temporal point processes, e.g., a multivariate spatial point process.
\end{example}

\begin{example}\label{ex3}
Let $\bXt=\bYt$ be the same stochastic point process. Then our proposed model includes the Hawkes process as a special case. The classical Hawkes process model only considers self-exciting effects, i.e., $\omega_{ij}(\cdot)$ has to be non-negative. Moreover, it usually assumes the process is stationary, which requires additional  conditions on $\omega_{ij}(\cdot)$, e.g., the spectral radius is smaller than one  \citep{bremaud1996}. We do not impose such constraints. 
\end{example}

\begin{example}\label{ex4}
For our motivating neuronal spike trains example, let $\bm{N}_{V1}(t)$, $\bm{N}_{A1}(t)$ denote the multivariate point process for the neuron firing activities on the layers V1 and A1, respectively.  If we expect that there are only directed connections between the neurons from A1 to V1 \citep{liang2013}, then we can set $\bYt=\bm{N}_{V1}(t)$, and $\bXt=\bm{N}_{A1}(t)$. Meanwhile, if we also expect potential connections between the neurons within the same layer of V1, then we can set $\bYt=\bm{N}_{V1}(t)$, and $\bXt=\left(\bm{N}_{V1}(t), \bm{N}_{A1}(t)\right)$, which takes the history of $\bm{N}_{V1}(t)$ into account as well.  Consequently, the transferring coefficient $\bm{\omega}_{A1}$ of $\bm{N}_{A1}(t)$ in this model captures the association between $\bm{N}_{V1}(t)$ and $\bm{N}_{A1}(t)$ after controlling the cross-connections within $\bm{N}_{V1}(t)$ itself. 
\end{example}

In our temporal point process modeling, we mainly target the conditional intensity function, instead of the moment statistics, for several reasons. To illustrate, we consider a special case of our model \eqref{eq:int_f}, which takes a linear link function, i.e., 
\begin{equation} \label{eq:linear}
\lambda^y_i(t)= \mu_i + \sum_{j=1}^p \left\{ \omega_{ij} * dX_{j} \right\} (t), \quad i=1,\ldots,m. 
\end{equation}
We next summarize the first and second-order statistics under this special case.  

\begin{prop} 
\label{thm:order}
Consider a special case of \eqref{eq:int_f}, such that $\bXt$ and $\bYt$ satisfy the linear relation \eqref{eq:linear}, and both are stationary. Then the corresponding moment statistics are of the form, 
\begin{equation} \label{eq:moments}
\begin{split}
& \bLambda^y  = \bmu + \left\{\int_{0}^{+\infty} \bomega(\Delta)d\Delta\right\} \; \bLambda^x, \quad 
\bC^{yx}(\tau) = \bomega(\tau) \textrm{Diag}(\bLambda^x) + \bomega * \bV^{xx} (\tau), \quad \tau \ge 0 \\
& \bV^{yy}(\tau) = \bomega * \bC^{xy}(\tau), \quad \tau>0, \quad \;\;
\bV^{yy}(0) = \bomega \star \left\{ \bV^{xx}(\cdot)+\textrm{Diag}(\bLambda^x) \right\}\star \bomega, 
\end{split}
\end{equation}
where $\bomega * \bC^{xy}(\tau) = \omega_{ij}(\cdot) * C_{ji}^{xy}(\tau)$, and $f * g(t) = \int f(\Delta) g(t-\Delta)d\Delta$ denotes the convolution of two univariate functions $f$ and $g$,  and $\bomega\star \big\{\bV^{xx}(\cdot)+\mbox{Diag}(\bLambda^x)\big\}\star \bomega=\int_{0}^{+\infty} \int _{0}^{+\infty} \bomega(\Delta)\big\{\textrm{Diag}(\bLambda^x)$ $\delta(\Delta'-\Delta)+\bV^{xx}(\Delta'-\Delta)\big\}\bomega\trans(\Delta') d\Delta d\Delta'$.  
\end{prop}
 
\noindent
The equations in \eqref{eq:moments} belong to a  class of integral equations for the Wiener-Hopf system with respect to $\bomega$. In principle, one can estimate the transferring function by solving the above equations and plugging in the estimated first and second-order statistics of $\bXt$ and $\bYt$. However, this strategy is not suitable for our model, because the moment-based estimation can be challenging when the dimension of the point process is high and diverging. Besides, the relation between the moment statistics and the transferring function in our model is much more complicated compared to a Hawkes process. For the latter, the second-order statistics full capture the associations between the marginal processes. In our model, however, the second-order statistics $\bV^{yy}$ and $\bC^{yx}$ also depend on the cross-covariance structure of the predictor process $\bXt$. Finally, nearly all moment-based estimation methods require the stationary condition, which can be restrictive in practice. For these reasons, we choose to adopt an intensity-based modeling approach.

\subsection{Low-rank structure}
\label{sec:lowrank}

In our model, the transferring function $\bomega$ that fully captures the cross-process connection pattern is of primary interest.  Adopting a common strategy in point process modeling, we assume that $\omega_{ij}(t)$ takes the form of a linear combination of a set of basis functions, $g^{(k)}(t), k=1, \ldots, K$, in that,
\begin{eqnarray} \label{eq:basis}
 \omega_{ij}(t) = \sum_{k=1}^K \beta_{ij}^k \cdot g^{(k)}(t), \quad i=1,\ldots,m, \; j=1,\ldots,p, 
\end{eqnarray}
where each $g^{(k)}(t)$ is a non-negative basis function on $[0, \infty)$,  $K$ is the number of basis functions, and $\beta_{ij}^k$'s can take positive and negative values.  The choice of basis functions mostly relies on the scientific knowledge to account for specific coevolutionary effects \citep{hansen2015}. In neuronal spike trains study, common basis functions include the exponential function, $g(t) = a\exp(-at), a>0$ \citep{zhou2013network}, the logarithmic decay function, $g(t)= \log(1+T-t)$, for the process defined on $[0,T]$ \citep{luo2016}, and the piecewise constant function, $g_{l}(t) = a_l \bm{1}(t \in \mathcal{T}_l)$, where $\{\mathcal{T}_l\}_{l=1}^{L}$ form a partition of $[0,+\infty]$ and $\{a_l\}_{l=1}^{L}$ are some non-negative constants \citep{wang2016}. One may also use a mix of different types of basis functions. 

Given the basis expansion in \eqref{eq:basis}, the conditional intensity model in (\ref{eq:int_f}) can be rewritten as, 
\begin{eqnarray} \label{eq:int_b}
\lambda_i^y(t) =\phi\left[ \mu_i+\sum_{j=1}^p \sum_{k=1}^K \beta_{ij}^{k} \cdot \left\{ g^{(k)} * dX_{j} \right\} (t) \right], \quad i=1,\ldots,m.
\end{eqnarray}
We then collect the transferring coefficients into a three-way tensor $\bcB \in \mR^{m \times p\times K}$, with the entry $\beta_{ij}^k$, $i=1, \ldots, m$, $j=1,\ldots,p$, $k=1,\ldots,K$. The conditional intensity function is now fully characterized by the background intensity vector $\bmu$ and the transferring coefficient tensor $\bcB$.  

We can estimate the model through a likelihood-based approach, where the joint log-likelihood function is of the form,  
\begin{eqnarray} 
\begin{split} \label{eq:loss0}
\cL\big(\bmu, \bcB|\bYt, \bXt \big) &=  \frac{1}{T}\sum_{i=1}^m L_{i}\big(\bmu, \bcB| Y_i(t), \bXt \big)\\
&=   \frac{1}{T}\sum_{i=1}^m \int_{0}^T \bigg[ \log\left\{ \lambda^y_i \big(t; \bcB, \bXt \big) \right\}dY_i(t) -\lambda^y_i \big(t; \bcB, \bXt \big)dt \bigg]. 
\end{split}
\end{eqnarray}
In the following discussion, we drop $\bYt$, $\bXt$ and $\bcB$ in $\cL(\bmu, \bcB)$ and $\lambda^y_i(t)$ for notational simplicity whenever there is no confusion. 

Next, we impose that $\bcB$ admits a low-rank CANDECOMP/PARAFAC (CP) structure, in that, 
\begin{eqnarray} \label{eq:cp}
\bcB=\sum_{r=1}^R \nu_r \bb^y_r \circ \bb^x_r \circ \bb^c_r, 
\end{eqnarray}
where $R$ is the tensor rank, $\bb^y_r\in \mR^{m}$, $\bb_r^x \in \mR^{p}$ and $\bb^c_r\in \mR^{K}$ are the normalized vectors corresponding to the modes of the response process, the predictor process, and the convolutional basis functions, respectively, $\nu_r$'s are the normalization weights, and ``$\circ$'' is the outer product. For notational convenience, we represent the decomposition \eqref{eq:cp} by a shorthand, $\bcB=\llbracket \bnu; \bB^{y}, \bB^{x},\bB^{c} \rrbracket$, where $\bB^y=[\bb_1^y \; \ldots \bb_R^y] \in \mR^{m\times R}$,  $\bB^x=[\bb_1^x \; \ldots \bb_R^x] \in \mR^{p\times R}$, $\bB^c=[\bb_1^c \; \ldots \bb_R^c] \in \mR^{K\times R}$, and $\bnu=(\nu_1, \ldots, \nu_R)\trans \in \mR^R$. See \citet{kolda2009} for a review of tensor and its decomposition. The low-rank decomposition \eqref{eq:cp} has been widely adopted in recent years in imaging-based tensor regressions \citep{zhou2013tensor, sun2017store, ChenYuan2019}. In the context of point process modeling, \cite{zhou2013network, Bacry2020} also adopted a  low-rank structure in a linear Hawkes model, yet on a matrix form of their transferring coefficients.  By contrast, we target a nonlinear, non-stationary, general temporal point process, and consider a low-rank structure on a coefficient tensor. Even though tensor is a conceptual generalization of matrix, tensor decomposition and matrix decomposition are considerably different \citep{kolda2009}. Naively transforming a tensor to a matrix may lose information. We also briefly remark that, there are easy-to-check sufficient conditions to ensure the decomposition in \eqref{eq:cp} is unique up to permutations \citep{sidiropoulos2000}. 

Imposing the low-rank structure like \eqref{eq:cp} in our point process regression has several advantages. First, it substantially reduces the number of free parameters in the transferring coefficient tensor $\bcB$, from $mpK$ to $R(m+p+K)$. In our example, if we set $\bYt=\bm{N}_{V1}(t)$, and $\bXt=\bm{N}_{A1}(t)$, the dimensions of the response and predictor processes are $m = 139$ and $p = 283$, respectively. If we choose $K = 3$ basis functions, and choose the rank $R = 4$,  then the number of free parameters in $\bcB$ reduces from $118,011$ to $1,700$. Second, and perhaps more importantly, it allows us to model the multivariate point process in a joint fashion. Existing approaches such as  \citet{hansen2015} model each response process $Y_i(t)$ separately, since their loss function is separable with respect to the individual intensity function $\lambda^y_i(t)$, each of which depends on a separate set of parameters $\bbeta_i$. By contrast, our model with \eqref{eq:cp} suggests that $\bcB$ relies on some underlying latent and interrelated factors, $\bB^y$, $\bB^x$ and $\bB^c$.  In the context of neuronal spike trains modeling, it implies that a particular predictor neuron in $\bXt$ exercises similar influence on multiple response neurons in $\bYt$, or a particular response neuron in $\bYt$ enjoys similar influence from multiple predictor neurons in $\bXt$. Unlike the separate modeling strategy, the latent factors  $\bB^x$ and $\bB^c$ are commonly shared by all intensity functions $\lambda_i(t)$'s, and thus information across different response processes $Y_i(t)$'s is integrated. Such an integration leads to an improved coefficient estimator, as we show asymptotically in Section \ref{sec:convergence}.

\subsection{Additional structure pursuit: sparsity and subgrouping}

To better accommodate scientific knowledge, improve the interpretation, and further reduce the number of free parameters, we consider some additional structure pursuit.  

The first structure we pursue is sparsity, in that only a subset of response processes are affected by a subset of predictor processes. This sparsity structure simplifies the model interpretation, further reduces the number of parameters, and is scientifically plausible. In multivariate Hawkes process modeling, the sparsity on transferring functions has been widely employed \citep{hansen2015, Bacry2020}. Specifically, we impose a group $\ell_1$ penalty \citep{yuan2006} on the coefficient tensor $\bcB$, 
\vspace{-0.1in}
\begin{eqnarray} \label{eq:sparsity}
P_{s}(\bcB; \tau_s) = \tau_s \sum_{i=1}^m \sum_{j=1}^p \left\| \bcB[i,j,\cdot] \right\|_2,
\end{eqnarray}
where $\bcB[i,j,\cdot] \in \mR^{R}$ is a vector of $\bcB$ with the first two indices fixed and the third index varying,  which corresponds to the associations between $Y_i(t)$ and $X_j(t)$ under all basis functions, $\tau_s$ is the sparsity tuning parameter, and $\|\cdot\|_2$ is the $\ell_2$ norm. 

The second structure we pursue is subgrouping. In a neuronal spike trains study, subsets of neurons are expected to share similar patterns in neuronal firing activities, and such clustering patterns are usually of great scientific interest \citep{kim2011granger}. Similar phenomenon is commonly observed in numerous other applications too. Such patterns are reflected by the underlying clustering structure in the transferring function $\bomega$. To capture this structure, we embed clustering pursuit into the proposed tensor decomposition. In principle, we can pursue clustering in the response process, or the predictor process, or both. For our motivating example, there is evidence of neuron clustering in the primary visual cortex V1, i.e., the response process \citep{liang2013}. As such, we introduce a grouping penalty on the decomposed factors of the response mode $\bB^y$, so to encourage clustering of the response neurons. Specifically, we impose a pairwise fusion penalty,  
\begin{eqnarray} \label{eq:fusion}
P_{f}(\bB^y; \tau_f) = \sum_{i < i'} f_{\kappa} \left( \left\| \bB^y[i,\cdot] - \bB^y[i',\cdot] \right\|_2, \tau_f \right), 
\end{eqnarray}
where $\bB^y[i,\cdot] \in \mR^{R}$ is the row vector of $\bB^y$, $\tau_f$ is the fusion parameter, and the penalty function $f_{\kappa}(t,\tau)=\tau\int_0^{t} \{ 1- x / (\tau \kappa) \}_+ dx$, with $\kappa$ being a thresholding parameter \citep{zhang2010}. This penalty function is to help reduce the estimation bias, as it only groups the individual predictors with similar effects on the responses through a non-convex fusion penalty \citep{zhu2019}.

\section{Estimation}
\label{sec:estimation}

\subsection{ADMM optimization}

We develop a highly scalable ADMM type optimization algorithm \citep{boyd2011} to estimate the parameters in our proposed model. Consider the realizations of the predictor and response processes $\bXt$ and $\bYt$ on a time interval $[0,T]$. Let $t^i_1 < t^i_2 < \cdots < t^i_{n_i}$ denote the time points of the $n_i$ events of the response process $Y_i(t)$ that are observed on $[0,T]$, $i=1, \ldots, m$.   Given the set of basis functions $\{g^{(k)}(\cdot)\}_{k=1}^K$, the  log-likelihood function in our model can be written as  
\begin{eqnarray} 
\cL(\bmu, \bcB)  =  \frac{1}{T} \sum_{i=1}^m \left( -\int_{0}^T \phi \left \{\mu_i+ \langle \bG (t), \bcB[i,\cdot,\cdot]\rangle \right\} dt  + \sum_{l=1}^{n_i} \log \left[ \phi \left \{\mu_i+ \langle \bG (t^i_l), \bcB[i,\cdot,\cdot]\rangle \right\} \right] \right), \nonumber
\end{eqnarray}
where $\bG(t)= \big( G_{j,k}(t) \big) \in \mR^{p \times K}$, $G_{j,k}(t)=\{ g^{(k)} * dX_{j} \} (t)$, $\bcB[i,\cdot,\cdot] \in \mR^{p \times K}$ is a matrix from $\bcB$ with the first index fixed and the other two indices varying, and $\langle \cdot, \cdot \rangle$ denotes the inner product. 

Incorporating the low-rank structure \eqref{eq:cp} and the two regularization structures \eqref{eq:sparsity} and \eqref{eq:fusion}, we aim at the following optimization problem, 
\vspace{-0.05in}
\begin{align} \label{eq:obj}
\begin{split}
\min_{\bmu, \bnu, \bB^{y}, \bB^{x},\bB^{c}} \; 
\Bigg\{ -\cL \big(\bmu, \llbracket \bnu; \bB^{y}, \bB^{x},\bB^{c} \rrbracket \big)
& + \tau_s \sum_{i=1}^m \sum_{j=1}^p \big\| \llbracket \bnu; \bB^{y}, \bB^{x},\bB^{c} \rrbracket [i,j,\cdot] \big\|_2 \\
& + \sum_{i < i'} f_{\kappa} \left( \big\| \bB^y[i,\cdot] - \bB^y[i',\cdot] \big\|_2, \tau_f \right) \Bigg\}.
\end{split}
\end{align}

\noindent 
The optimization in (\ref{eq:obj}) is challenging in several ways. It involves a tensor decomposition embedded in a complicated log-likelihood function with summation of integrals and a possibly nonlinear link function $\phi$. In addition, the sparsity penalty in (\ref{eq:sparsity}) is non-differentiable, while the fusion penalty in (\ref{eq:fusion}) is non-convex. Moreover, (\ref{eq:fusion}) involves the differences of parameters, rendering those parameters inseparable in optimization. To overcome those challenges, and to achieve computational scalability, we develop an ADMM algorithm for the optimization in (\ref{eq:obj}). 

Specifically, we introduce three sets of auxiliary variables. The first set is $\bcA \in \mR^{m \times p \times K}$ that targets the low-rank structure \eqref{eq:cp} such that $\bcA = \bcB = \llbracket \bnu; \bB^{y}, \bB^{x},\bB^{c} \rrbracket$. The second set is $\bPsi \in \mR^{m \times p \times K}$ with $\bPsi[i,j,\cdot] = \bpsi_{ij} \in \mR^{K}$ that targets the sparsity structure \eqref{eq:sparsity} such that $\bpsi_{ij} = \bcB[i,j,\cdot]$, $1 \le i \le m, 1 \le j \le p$. The third set is $\bGamma \in \mR^{m(m-1)/2 \times R}$ that stacks $\bm{\gamma}_{ii'} \in \mR^{R}$ together and targets the subgroup structure \eqref{eq:fusion} such that $\bm{\gamma}_{ii'} = \bB^y[i,\cdot]-\bB^y[i',\cdot]$, $1 \le i < i' \le m$. We then rewrite (\ref{eq:obj}) in its equivalent form, 
\begin{align} \label{eq:copt}
\begin{split}
\min_{\bmu, \bnu, \bB^{y}, \bB^{x},\bB^{c}, \bcA, \bPsi, \bGamma} & \left\{ -\cL(\bmu, \bcA) + \tau_s \sum_{i=1}^m \sum_{j=1}^p \|\bm{\psi}_{ij}\|_2 + \sum_{j<j'}f_{\kappa}\left( \|\bm{\gamma}_{ii'}\|_2, \tau_f \right) \right\} \\
& \; \textrm{ subject to } \; \bcA = \llbracket \bnu; \bB^{y}, \bB^{x},\bB^{c} \rrbracket, \quad  \bPsi = \bcB,  \quad \bGamma = \bD_m \bB^y,
\end{split}
\end{align}
where $\bD_m \in \mR^{m(m-1)/2 \times m}$ that stacks $\bd_{ii'} \in \mR^{m}$ together, with $\bd_{ii'} = \be_i - \be_{i'}$, $\be_i \in \mR^{m}$ has one on the $i$th position and zero elsewhere, $1 \le i < i' \le m$. Note that the second constraint $\bPsi = \bcB$ can also be written as $\bPsi = \bcA$ due to the first constraint. To solve \eqref{eq:copt}, we minimize the following augmented Lagrangian objective function, 
\vspace{-0.05in}
\begin{eqnarray*}
& & -\cL(\bcA, \; \bmu)+\tau_s \sum_{i,j} \|\bm{\psi}_{ij}\|_2+\sum_{j<j'}f_{\kappa}(\|\bm{\gamma}_{ii'}\|_2, \tau_f) \\
& & + \; \langle \bcW_1, \bcA - \llbracket \bnu; \bB^{y}, \bB^{x},\bB^{c} \rrbracket \rangle + \langle \bcW_2, \bm{\Psi} - \bcA \rangle + \langle \bW_3, \bGamma - \bD_m \bB^y \rangle \\
& & + \; \frac{\rho}{2} \left( \left\| \bcA - \llbracket \bnu; \bB^{y}, \bB^{x},\bB^{c} \rrbracket \right\|_F^2 + \| \bm{\Psi} - \bcA \|_F^2+ \| \bGamma - \bD_m \bB^y \|_F^2 \right),
\end{eqnarray*}
where $\bcW_1, \bcW_2 \in \mR^{m\times p\times K}$ and $\bW_3 \in \mR^{m(m-1)/2 \times m}$ are the corresponding Lagrangian multipliers, $\rho >0$ is a fixed augmented parameter, and $\|\cdot\|_F$ denotes the Frobenius norm.  

Next we update the blocks of parameters, $\bmu, \bcA, \bnu, \bB^{y}, \bB^{x},\bB^{c}, \bPsi, \bGamma$, and the Lagrangian multipliers $\bcW_1, \bcW_2, \bW_3$ in an alternating fashion. That is, given the estimates at the $s$th iteration, $\bcA^{(s)}, \bcB^{(s)}= \llbracket \{\bnu\}^{(s)}; \{\bB^{y}\}^{(s)}, \{\bB^{x}\}^{(s)},\{\bB^{c}\}^{(s)} \rrbracket, \bPsi^{(s)}, \bGamma^{(s)}, \bcW_1^{(s)}, \bcW_2^{(s)}, \bW_3^{(s)}$, we update:
\begin{align}
\begin{split} \label{eq:admm1}
\bmu^{(s+1)}, \bcA^{(s+1)} & =  \argmin_{\bmu, \bcA}  -\cL(\bcA, \; \bmu) + \frac{\rho}{2}\left\{ \left\| \bcA - \bcB^{(s)} + \rho^{-1} \bcW_1^{(s)} \right\|_F^2 \right.  \\
 & \hspace{2.3in} \left. + \left\| \bcA - \bm{\Psi}^{(s)} + \rho^{-1} \bcW_2^{(s)} \right\|_F^2\right\}, 
\end{split} \\
\begin{split} \label{eq:admm2}
\bcB^{(s+1)} & =  \argmin_{\bnu, \bB^{y}, \bB^{x},\bB^{c}} \left\| \cp-\bcA^{(s+1)}-\rho^{-1} \bcW_1^{(s)} \right\|_F^2 \\
 & \hspace{2.3in} + \left\| \bGamma^{(s)} - \bm{D}_p \bB^{y} - \rho^{-1} \bW_3^{(s)} \right\|_F^2, 
\end{split} \\
\bm{\Psi}^{(s+1)} & = \argmin_{\bm{\Psi}} \frac{\rho}{2}\left\|\bm{\Psi}-\bcA^{(s+1)}-\rho^{-1} {\bcW}_2^{(s)} \right\|_F^2 + \tau_s\sum_{i ,j}\|\bm{\psi}_{ij}\|_2, \label{eq:admm3} \\
\bGamma^{(s+1)} & =  \argmin_{\bGamma} \frac{\rho}{2} \left\|\bGamma-\bm{D}_p (\bB^y)^{(s+1)}-\rho^{-1} \bm{W}_3^{(s)} \right\|_F^2 + \sum_{j<j'}f_{ \rho}(\|\bm{\gamma}_{ii'}\|_2, \tau_f). \label{eq:admm4} \\
\begin{split} \label{eq:admm5}
\bcW_1^{(s+1)} & =  \bcW_1^{(s)} + \rho \left\{ \bcA^{(s+1)} - \bcB^{(s+1)} \right\}, \\
\bcW_2^{(s+1)} & =  \bcW_2^{(s)} + \rho \left\{ \bcA^{(s+1)} - \bm{\Psi}^{(s+1)} \right\},  \\
\bW_3^{(s+1)} & =  \bW_3^{(s)} + \rho \left\{ \bD_m (\bB^y)^{(s+1)}-\bGamma^{(s+1)} \right\}. 
\end{split}
\end{align}
We then tackle the optimization problems (\ref{eq:admm1}) to (\ref{eq:admm4}) one-by-one. 

The optimization problem in (\ref{eq:admm1}) can be split slice-by-slice for $\bcA[i,\cdot,\cdot]$, $i=1,\ldots,m$. That is, it can be solved with respect to each marginal response process $Y_i(t)$ in a parallel fashion. Define 
\vspace{-0.05in}
\begin{align*} \label{eq:admm11}
\begin{split}
& L_i^*(\mu_i, \bcA[i,\cdot,\cdot]) =  \sum_{l=1}^{n_i} \log \left[ \phi \left\{ \mu_i+ \langle \bG (t^i_l), \bcA[i,\cdot,\cdot]\rangle \right\} \right]  -\int_{0}^T \phi \left\{\mu_i+ \langle \bG (t), \bcA[i,\cdot,\cdot]\rangle \right\} dt  \\
& + \; \frac{\rho}{2}\left\|\bcA{[i,\cdot,\cdot]} - \bcB^{(s)}{[i,\cdot,\cdot]}+\rho^{-1} \bcW_1^{(s)}{[i,\cdot,\cdot]}\right\|_F^2 
+ \frac{\rho}{2}\left\|\bcA{[i,\cdot,\cdot]} - \bm{\Psi}^{(s)}{[i,\cdot,\cdot]}+\rho^{-1} \bcW_2^{(s)}{[i,\cdot,\cdot]}\right\|_F^2. 
\end{split}
\end{align*}
The objective function $L_i^*(\mu_i, \bcA[i,\cdot,\cdot])$ is differentiable, and with a large enough $\rho$, it is almost convex regardless of the form of the link function $\phi$. Therefore, we can minimize $L_i^*(\mu_i, \bcA[i,\cdot,\cdot])$ efficiently using a gradient descent type algorithm. In our implementation, we employ the Newton-Raphson algorithm, as we use the linear and the logit link functions. 

The optimization problem in (\ref{eq:admm2}) turns to be a regularized CP decomposition with an $\ell_2$ penalty. It can be solved by an alternating block updating algorithm \citep{zhou2013tensor}, which updates one block of the parameters in $\{\bB^y, \bB^x, \bB^c\}$, while fixing the other two blocks and $\bnu$. For instance, $\bB^y$ is updated by minimizing $\Big\| \left\{ \bcA^{(s+1)}+\rho^{-1} \bcW_1^{(s)} \right\}_{(1)} - \bB^y \ \left[(\bB^c)^{(s)} \odot (\bB^x)^{(s)}\;\mathrm{diag}\left\{ \bnu^{(s)} \right\} \right]\trans \Big\|^2$ $+ \Big\| \bGamma^{(s)}  - \rho^{-1} \bW_3^{(s)} -\bm{D}_m \bB^{y} \Big\|^2$, with respect to $\bB^y$, where $\odot$ is the Khatri-Rao product, $\bcA_{(1)}$ denotes the mode-$1$ matricization of the tensor $\bcA$, and $\mathrm{diag}(\bnu)$ is the diagonal matrix with $\bnu$ as the diagonal elements. Note that this is essentially a least squares optimization problem with an $\ell_2$ penalty, which has an explicit solution. The other two blocks $\bB^x$ and $\bB^c$ are updated similarly. After updating each block, for instance, $\bB^y$, we update $\nu_r$ by normalizing the corresponding vector $\bb^y_r$, respectively. In addition, we employ the maximum block improvement strategy of  \cite{chen2012} to ensure the convergence of the alternating block updating iterations \citep{tang2019}.

The optimization problems in \eqref{eq:admm3} and \eqref{eq:admm4} have explicit solutions, since the corresponding objective functions are convex with respect to $\bm{\psi}_{ij}$, and $\bm{\gamma}_{ii'}$ when $\kappa>\rho^{-1}$, respectively. That is, 
\begin{eqnarray} 
\bm{\psi}^{(s+1)}_{ij} & = & \left\{ 
\begin{array}{lcl}
\bm{0} & \quad \text{if} \; \|  \bvartheta_{ij}^{(s+1)}\| < \sqrt{K}\tau_s/\rho, \\
\left\{ 1- \frac{\sqrt{K}\tau_s/\rho}{\|\bvartheta^{(s+1)}_{ij}\|} \right\} \bvartheta^{(s+1)}_{ij} & \quad \text{if} \; \|  \bvartheta_{ij'}^{(s+1)}\| \ge \sqrt{K}\tau_s/\rho,
\end{array} \right. \label{eq:admm41} \\
\bm{\gamma}^{(s+1)}_{ii'} & = & \left\{
\begin{array}{lcl}
\bzeta_{ii'}^{(s+1)} & \quad \text{if} \; \|  \bzeta_{ii'}^{(s+1)}\| \ge \kappa \tau_f, \\
\frac{\kappa\rho}{\kappa\rho-1}\left\{ 1- \frac{\tau_f/\rho}{\|\bm{\zeta}^{(s+1)}_{ii'}\|} \right\}_{+} \bzeta^{(s+1)}_{ii'} & \quad \text{if} \; \|  \bzeta_{ii'}^{(s+1)}\| < \kappa \tau_f,
\end{array} \right. \label{eq:admm31}
\end{eqnarray} 
where $\bvartheta_{ij}^{(s+1)}=\bcA[i,j,\cdot]^{(s+1)} +\rho^{-1}\{\bcW_2[i,j,\cdot]\}^{(s)}$, $\bzeta_{ii'}^{(s+1)} = {(\bB^y[i,\cdot])}^{(s+1)}- {(\bB^y[i',\cdot])}^{(s+1)} +\rho^{-1}\bm{W}_3[l_{ii'},\cdot]^{(s)}$, and $l_{ii'}=(2m-i)(i-1)/2+i'-i$. We note that this computation can be done in a parallel fashion over $(i, i'), i, i' = 1, \ldots, m$, and $(i, j), i=1,\ldots,m, j=1,\ldots,p$.

We summarize the above optimization procedures in Algorithm \ref{alg}. 

\begin{algorithm}[t!]
\caption{The ADMM algorithm for parameter estimation.}
\label{alg}
\begin{algorithmic}
\STATE [1] Initialize $\bmu^{(0)}, \cp^{(0)}, \bcA^{(0)}, \bm{\Psi}^{(0)}, \bGamma^{(0)}, \bcW_1^{(0)}, \bcW_2^{(0)}, \bm{W}_3^{(0)}$.  Set $\rho$ and $\kappa>\rho^{-1}$. 
\REPEAT
\STATE [2] Update $\bmu_i^{(s+1)}, \bcA{[i,\cdot,\cdot]}^{(s+1)}$ via (\ref{eq:admm1}) with parallel computing over $i=1,\ldots,m$.
\STATE [3] Update  $\left\llbracket \bnu^{(s+1)}; {\{\bB^y\}}^{(s+1)}, {\{\bB^x\}}^{(s+1)}, {\{\bB^c\}}^{(s+1)}\right\rrbracket$ via (\ref{eq:admm2}).
\STATE [4] Update  ${\bm{\Psi}}^{(s+1)}=\left\{\bm{\psi}^{(s+1)}_{ij}\right\}$ via (\ref{eq:admm41}) with parallel computing over $1 \le i \le m, 1\le j \le p$.
\STATE [5] Update  $\bGamma^{(s+1)}=\left\{ \bm{\gamma}^{(s+1)}_{ii'} \right\}$ via (\ref{eq:admm31}) with parallel computing over $1 \le i < i' \le m$.
\STATE [6] Update $\bcW_1^{(s+1)}, \bcW_2^{(s+1)}, \bm{W}_3^{(s+1)}$ via (\ref{eq:admm5}).
\UNTIL{the stopping criterion is met.}
\end{algorithmic}
\end{algorithm}

\subsection{Initialization, convergence, tuning, and computational complexity}
\label{sec:tuning}

We recommend to use a warm initialization, by setting the initial values $\{\bmu^{(0)}, \bcB^{(0)}\}$ as the unpenalized estimators without imposing any low-rank or penalty structures, while setting the other initial values at zeros.  We stop the algorithm when some stopping criterion is met, e.g., when the difference of the consecutive estimates is smaller than a threshold. 

Algorithm \ref{alg} converges to a stationary point.  This can be verified by checking the conditions of Proposition 1 in \cite{zhu2019}. 
\begin{prop}
Suppose the log-likelihood function $\cL$ is a Lipschitz function with respect to $\bcB$, and the parameter space for $\bB^y$, $\bB^x$ and $\bB^c$ is a compact set. Then the obtained estimator from Algorithm \ref{alg} converges to a stationary point of the objective function in \eqref{eq:obj}.
\end{prop}

We select the tuning parameters as follows. The first is the Lagrangian augmented parameter $\rho$, which can be viewed as the learning rate of the ADMM algorithm. Our numerical results have suggested that the final estimates are not overly sensitive to the choice of $\rho$, so we simply set $\rho = 1$. The second is the thresholding parameter $\kappa$ in the fusion penalty $f_\kappa$. Again, the estimates are not sensitive to $\kappa$ as long as $\kappa > \rho^{-1}$, and we set $\kappa = 2$. The third set of tuning parameters include the rank $R$ in  \eqref{eq:cp}, and the two regularization parameters, $\tau_s$ in \eqref{eq:sparsity} and $\tau_f$ in \eqref{eq:fusion}. We tune them by minimizing a Bayesian information criterion (BIC), $-2\mathcal{L}(\bmu, \bcB) + \log(N)p_e$, where $N=\sum_{i=1}^{m} n_i$ is the total number of events observed on the multivariate response process $\bYt$, and $p_e$ is the effective number of parameters. For the tuning of $R$, $p_e = R(m+p+K-2)$, for $\tau_s$, $p_e$ is the total number of  non-zero latent parameters, and for $\tau_f$, $p_e$ is the total number of unique non-zero latent parameters. A similar BIC type criterion has been commonly adopted in low-rank tensor regressions \citep{zhou2013tensor, sun2017store}. Moreover, to speed up tuning, we tune $R, \tau_s, \tau_f$ in a sequential manner. That is, we first tune $R$ while setting $\tau_s = \tau_f = 0$, then tune $\tau_s$ given the selected $R$ while setting $\tau_f=0$, and finally tune $\tau_f$ given the selected $R$ and $\tau_s$.   

Finally, we briefly discuss the computational complexity of Algorithm \ref{alg}. As an example, if we use a sigmoid link function, then the overall computational complexity can be approximated by $O\left[n_{iter} \left\{ mC_{LR(p)} + C_{ALS(mp)} + mp+m(m-1)/2 \right\} \right]$, where  $n_{iter}$ denotes the total number of ADMM iterations,  $C_{LR(p)}$ denotes the computational complexity for a logistic regression with $p$ covariates, and $C_{ALS(mpK)}$ denotes the computational complexity for a three-way tensor decomposition with the size $m\times p\times K$ using the alternating least squares (ALS) algorithm. Besides, several steps in this algorithm can be accelerated using parallel computing. In Section \ref{sec:simulations}, we report the computation time of our simulated examples.

\section{Theory}
\label{sec:theory}

\subsection{Regularity conditions}
\label{sec:conditions}

We begin by introducing some notation. Let $\bth=\big\{ \bmu\trans, \vec(\bB^y)\trans,\vec(\bB^x)\trans,\vec(\bB^c) \big\}\trans$ collect all latent parameters in our model, including the background intensity $\bmu$, and the latent factors $\bB^y, \bB^x, \bB^c$ from the CP decomposition \eqref{eq:cp}. Without loss of generality, the normalization weight $\bm{\nu}$  is omitted to simplify the notation. Let $\bbeta = \bbeta(\bth) = \left(\bbeta_1\trans, \ldots, \bbeta_m\trans \right)\trans$, where $\bbeta_i = \big\{ \mu_i, \vec\big( \bcB[i,\cdot,\cdot] \big)\trans \big\}\trans$, $i=1,\ldots,m$. Note that the transferring coefficient $\bbeta$ is a function of $\bth$, and thus we sometimes write it as $\bbeta(\bth)$. Let $\bTh_{\bth} \subset \mR^{R(m+p+K)+m}$ and $\bTh_{\beta} \subset \mR^{mpK+m}$ denote the parameter space for $\bth$ and $\bm{\beta}$, respectively.  For a real-valued function $f(t)$ defined on $[0,\infty)$, define the norm $\|f\|_A = \big\{ \int_{A} f^2(t)dt \big\}^{1/2}$, where $A$ is a Borel set  in $[0,\infty)$. In particular, write $\|f\|_T = \big\{ \int_{0}^T f^2(t)dt \big\}^{1/2}$ for any $T>0$.  Moreover, let $\|\cdot\|_2$, $\|\cdot\|_{\infty}$, $\|\cdot\|_F$, and $\|\cdot\|_{max}$ denote the $\ell_2$ norm, the $\ell_{\infty}$ norm,  the Frobenius norm, and the maximum norm, respectively. Let $\pi_{min}(\cdot)$ and $\pi_{max}(\cdot)$ denote the smallest and the largest eigenvalue for a symmetric matrix. 

For our theoretical analysis, we consider a general likelihood-based loss function, which encompass our model \eqref{eq:int_f} and the two penalty functions \eqref{eq:sparsity} and \eqref{eq:fusion}, 
\begin{eqnarray} \label{eq:loss}
\mc{S}\{\bbeta(\bth)\} = -\cL\{\bbeta(\bth)\}  + \tau \, P\{\bbeta(\bth)\}  = -\frac{1}{T}\sum_{i=1}^m L_{i}\{\bbeta(\bth)\}  + \tau \, P(\bth), 
\end{eqnarray}
where $L_i(\cdot)$ is the log-likelihood function for the $i$th response process $Y_i(t)$, while in our model it is as specified in \eqref{eq:loss0}, $i=1, \ldots, m$, $P(\cdot)$ is a non-negative penalty function, and $\tau$ is the penalization parameter. Note that $\bth$ is associated with the log-likelihood function $L_i(\cdot)$ only through $\bbeta(\bth)$. 

We next present a set of regularity conditions, where $c_1$ to $c_5$ are some finite positive constants. 

\vspace{-0.05in}
\begin{enumerate}[({C}1)]
\item  Let $\cK(\bbeta^0) \subset \bTh_{\bbeta}$ denote a neighborhood of  the true value $\bbeta^0$.  For any $\bbeta, \tilde{\bbeta} \in \cK(\bbeta^0)$ and a large enough $T$, $\lambda_{i}^y(t; \bbeta) = \lambda_{i}^y(t; \tilde{\bbeta})$ almost surely on $[0,T]$, if and only if $\bbeta = \tilde{\bbeta}$. 

\item  For any $i = 1, \ldots, m$, $\sup_{\mc{V}(A)<1} \mE\{Y_i(A)\}^2 / \mc{V}(A) < \infty$, and for any $j = 1, \ldots, p$, $\sup_{\mc{V}(A)<1}$ $\mE\{X_j(A)\}^2 / \mc{V}(A) < \infty$, where $A$ is a Borel-set on $[0, +\infty)$, and $\mc{V}$ is the Lebesgue measure. 

\item For any  $\bbeta \in \cK(\bbeta^0)$, $i = 1, \ldots, m$, \hspace{0.5em}$\sup_{t \in [0,T]}\mE\left[\left\|\frac{\partial \log\{\lambda_i^y(t)\}}{\partial \bbeta_i} \lambda^y_i(t; \bbeta^0)\right\|_\infty^2\right] < \infty$,  \\$\sup_{t \in [0,T]}\mE\left[\left\|\frac{\partial^2 \log\{\lambda_i^y(t)\}}{\partial \bbeta_i\partial \bbeta_i\trans} \lambda^y_i(t; \bbeta^0)\right\|_{max}^2 \right] < \infty$, and  $\sup_{t \in [0,T]}\mE\left\{ {\lambda_i^y(t)^2}\right\} < \infty$.

\item For any $\bbeta \in \cK(\bbeta^0)$, $i = 1, \ldots, m$, $\pi_{min} \left( \bJ_{i}\right) \ge c_1$ almost surely with a large $T$, where $\bJ_{i} = T^{-1}\int_{0}^T \bH_{i}(t) \bH_{i}(t)\trans\lambda^y_i(t,\bbeta_i^0) dt$, and $\bH_i(t) = \lambda_i^y(t)^{-1} \partial \lambda^y_i(t) / \partial \bbeta_i$.

\item The link function $\phi(x)$ is a Lipschitz function satisfying that $|\phi(x_1)-\phi(x_2)| \le c_2 |x_1-x_2|$ for any  $x_1, x_2 \in \mR$ and some $c_2$. 

\item The basis function $g^{(k)}(t)$ satisfies that $\max_{1\le k \le K} \left\{\int_0^{\infty} g^{(k)}(t)^2dt\right\}^{1/2}  < c_3$ for some $c_3$.

\item The penalty function $P(\bth)$ is a non-negative Lipschitz function in a neighborhood of the true value $\bth^0$ satisfying that $|P(\bth_1)-P(\bth_2)| \le c_4 \|\bth_1-\bth_2\|_2$ for some $c_4$. 

\item Let $\mc{I}_1,\ldots, \mc{I}_N$ denote the true subgroup partition of the index set $\{1,\ldots,m\}$, in that $\bB^y[i, \cdot]$ $= \bar{\bb}_{(s)}$ for any $i \in \mc{I}_s$, $s = 1, \ldots, N$, and $N$ is the number of subgroups. There is a minimum gap, such that $\min_{s\neq s'} \left\| \bar{\bb}_{(s)} -\bar{\bb}_{(s')}\right\|_2 > c_5$.
\vspace{-0.05in}
\end{enumerate}

\noindent
We make some remarks about these conditions. Condition (C1) ensures the identifiability of the intensity function with respect to $\bbeta$, and is equivalent to \citet[Assumption B3]{ogata1978}. Condition (C2) implies a finite mean intensity for both the response and the predictor process, and the resulting process is referred as a non-explosive point process \citep{daley2007}. The same condition was imposed in \citet[Assumption A3]{ogata1978} and \citet[condition of Theorem 2]{hansen2015}. Such a condition also makes sense in neuronal activity studies, which implies that there is an upper bound for the average neuronal activity level or calcium concentration level over time.  Condition (C3) is a standard regularity condition, and the same condition or its equivalent forms have been commonly adopted in the point process literature; see, e.g., \citet[Conditions B4, B5]{ogata1978} for a stationary process, and \citet[Conditions C1, C4]{rathbun1994} for an inhomogeneous process. This condition is also easy to verify for a class of commonly used link functions, e.g., a rectifier link or a sigmoid link.  Condition (C4) is placed on the minimum eigenvalue of the information matrix. A similar condition was considered in \citet[Assumption B6]{ogata1978} and \citet[Condition c]{rathbun1994}, and it is analogous to the usual regularity condition placed on the design matrix for regressions with random variables.  Condition (C5) is usually adopted  in nonlinear point process models \citep{bremaud1996}, and it holds for a range of commonly used link functions.  Similarly, Condition (C6) holds for numerous basis functions, since the basis function is generally a normalized decaying kernel. Condition (C7)  holds for a variety of penalty functions in a compact space, including the $\ell_1$ and $\ell_2$ penalties, the group $\ell_1$ penalty in \eqref{eq:sparsity}, and the fusion penalty in \eqref{eq:fusion}. Finally, Condition (C8) is a standard condition to ensure the identifiability of the subgroups, and has been often assumed in subgroup analysis \citep{ma2017subgroup, zhu2019}. This condition is not required for establishing the coefficient estimation convergence properties, but only for the subgroup identification consistency. In summary, we feel the above regularity conditions are relatively mild and reasonable. They are clearly weaker than the stationary condition. The same conditions or similar forms have been widely adopted in the asymptotic studies of temporal point process in the literature.

\subsection{Convergence properties}
\label{sec:convergence}

We next derive the asymptotic properties for the penalized likelihood estimator from \eqref{eq:loss}, which covers a variety of link and penalty functions, and does not assume the stationarity. We allow the point process dimension to diverge, and show that the increasing dimension is to actually benefit the estimation, leading to a faster convergence rate for the coefficient estimator and a smaller error bound for the recovered intensity function. In the interest of space, we present some supporting lemmas and the proofs in the Supplementary Appendix. 

Since the parameter spaces $\bTh_{\bth}$ and $\bTh_{\beta}$ grow along with the point process dimensions, we adopt the sieve idea and the large-deviation approach introduced by \citet{shen1994, shen1998} to derive the asymptotics.  Specifically, we define a restricted parameter space for $\bth$, 
\vspace{-0.05in}
\begin{eqnarray*}
\tilde{\bTh}_{\bth} = \left\{ \bth \in \bTh_{\bth}: \|\bth\|_{\infty} \le c_{0}, P(\bth) \le c_{\bth}^2 \right\},
\end{eqnarray*}
where $c_{0}$ is a positive constant, and $c_{\bth}$ is another constant that is allowed to increase at the rate of $O\left(\sqrt{(m+p+K)R+m}\right)$, since the dimension of $\bth$ is $(m+p+K)R+m$ that is to diverge with $m$ and $p$. Furthermore, we define a metric based on the Kullback-Leibler (KL) pseudo-distance in $\bTh_{\bth}$ with respect to the underlying true value $\bth^0$ as, 
\begin{eqnarray*}
d(\bm{\theta}, \bm{\theta}^0) = \frac{1}{\sqrt{mp}}\mE\left[ \cL\{\bbeta(\bth)\} - \cL\{\bbeta(\bth^0)\} \right]^{1/2}.
\end{eqnarray*}
Analogous to  \citet[Lemma 3]{ogata1978},  it is straightforward to verify that $d(\bth, \bth^0)$ is an appropriate  distance metric for any $\bth \in \bTh_{\bth}$.  Let $\hat{\bth}=\argmin_{\bth \in \tilde{\bTh}_{\bth}} \mc{S}\{\bbeta(\bth)\} $ denote the penalized likelihood estimator for \eqref{eq:loss}. The next theorem shows that $\hat{\bth}$ converges to the true value $\bth^0$ exponentially in probability under the KL distance. 
 
\begin{thm} \label{thm:KL}
Suppose Conditions (C1) to (C7) hold. For some $\varepsilon_1 > 0$, there exist finite positive constants $\tc_1, \tc_2$, such that 
\vspace{-0.05in}
\begin{eqnarray*}
\P\left\{ d(\hat{\bth}, \bth^0)  \ge \varepsilon_1 \right\} \leq 7\exp \left(-\tc_2  T \eta^2_{\bth} \varepsilon_1^2\right),
\end{eqnarray*}
where 
\vspace{-0.1in}
\begin{eqnarray*}
\eta_{\bth}=\frac{(mpK)^{1/2}}{\{R(m+p+K)+m\}^{1/2}} \left[\log\left\{ \frac{\tc_1 mpK}{\sqrt{R(m+p+K)+m}}\right\}\right]^{-1/2}, 
\end{eqnarray*}
 and the penalty parameter $\tau$ in \eqref{eq:loss} satisfies that $\tau \le  O(T^{-1} \eta^{-2}_{\bth})$.
\end{thm}

The convergence result in Theorem \ref{thm:KL} is established under the KL distance, which is stronger than and usually dominates some other distance measures, e.g., the Hellinger metric. The next corollary establishes the convergence of the recovered transferring coefficient $\bcB$ under the $\ell_2$ norm. Denote $\hat{\bcB}=\bcB(\hat{\bth})$, $\bcB^0 = \bcB(\bth^0)$, and $\tilde{d}(\hat{\bcB},\bcB^0) = (mpK)^{-1/2} \|\hat{\bcB} -\bcB^0\|_F$. 

\begin{corollary} \label{coro:L2}
Suppose the conditions in Theorem \ref{thm:KL} hold.  For some $\varepsilon_2 >0$, there exists a finite positive constant $\tc_3$, such that 
\vspace{-0.05in}
\begin{eqnarray*}
\P\left\{ \tilde{d}(\hat{\bcB},\bcB^0) \ge \varepsilon_2 \right\} \leq 7\exp \left(-\tc_3  T\eta_{\bcB}^2 \varepsilon^2_2\right),
\end{eqnarray*}
where $\eta_{\bcB} = [(mpK) / \{R(m+p+K)+m\}]^{1/2}$. 
\end{corollary}

A few remarks are in order. First, Theorem \ref{thm:KL} and Corollary \ref{coro:L2} indicate that the penalized estimator and the recovered coefficient tensor achieve a convergence rate of $\sqrt{T}\eta_{\bth}$ and $\sqrt{T}\eta_{\bcB}$, respectively.  On one hand, since the length $T$ of the observed point process plays the role of sample size as in the usual random variable based regressions, an increasing $T$ would lead to a faster convergence rate and a smaller error bound. On the other hand, the diverging point process dimension $m$ and $p$ are to benefit the estimation as well. This is achieved not because of a stronger set of model and regularity conditions we impose; actually as we discuss in detail earlier that our conditions are compatible with or weaker than those in the existing literature. But it is indeed due to our proposed low-rank model structure  \eqref{eq:cp}. Specifically,  our model substantially reduces the size of the parameter space $\bTh_{\bbeta}$ through $\bcB(\bth)$ with the latent parameter $\bth \in \bTh_{\bth}$.  Consequently, it enables us to obtain a smaller metric entropy with bracketing on the restricted parameter space $\tilde{\bTh}_{\bth}$, which in turn yields a tighter bound for the loss function and a faster convergence rate. Moreover, the latent factors $\bB^x$ and $\bB^c$ are commonly shared across the intensity functions of all response processes. This enables us to utilize and borrow information from the entire multivariate response process $\bYt$ rather than a single response $Y_i(t)$. This result of the blessing of dimensionality clearly distinguishes our method from the existing ones; e.g., \citet{Bacry2020} studied the asymptotics for a multivariate Hawkes process model and potentially allowed the process dimension to diverge, but their error bound is to increase at the rate of the logarithm of the dimension when it diverges.  Second, in our current analysis, we fix the rank $R$ of the tensor decomposition \eqref{eq:cp} for simplicity. However, we can allow $R$ to diverge along with $m$ and $p$ too. By Corollary  \ref{coro:L2}, as long as $R$ grows at a limited rate of $o\left(\min(m,p)\right)$, e.g., $\log(m)$ or $\log(p)$, the obtained estimator still enjoys a faster convergence rate as $m$ and $p$ increase. Third, we note that the theoretical properties obtained in Theorem \ref{thm:KL} and Corollary \ref{coro:L2} are for the global minimizer of \eqref{eq:loss}. Nevertheless, \eqref{eq:loss} is a non-convex optimization problem, and there is no guarantee that the optimization algorithm can land at the global minimizer. This is a well-known issue in almost any statistical models involving non-convex optimization \citep{zhu2016}, and it still remains an open question. In recent years, there has been some progress to tackle this problem. For instance, \citet{bi2018} showed that, in a tensor factorization model, the established large deviation property of a global minimizer can be generalized to an asymptotically good local optimizer. However, it was obtained with the price of imposing additional assumptions. We leave this problem as future research. 

Finally, we establish the subgroup structure identification consistency based on the estimated transferring coefficients.  

\begin{thm} \label{thm:subgroup}
Suppose Conditions  (C1) to (C8) hold. Let $\hat{\bB}^y$ denote the estimated latent factor $\bB^y$ from \eqref{eq:obj}. Suppose $\tau_s=o\left\{ \left(T\eta_{\bcB}^2\right)^{-1/2}\right\}$, and $\tau_f=O\left\{\left(T\eta_{\bcB}^2\right)^{-1/2+c_f}\right\}$ for $0<c_f<1/2$. Then, 
\begin{eqnarray*}
\P\left( \hat{\bB}^y[i,\cdot] = \hat{\bB}^y[i',\cdot] \; |\; i, i' \in \mc{I}_s, \;1\le s \le N \right) \rightarrow 1, \;\; \mathrm{ as } \;\; T \rightarrow \infty.
\end{eqnarray*}
\end{thm}

Theorem \ref{thm:subgroup} shows that, as $T \rightarrow \infty$, the true subgroup structure can be identified with the probability tending to one.  We also comment that,  the number of subgroups is also allowed to increase  as $m$ increases, as long as Condition (C8) holds. Moreover, we may relax (C8), by allowing the minimum gap $c_5$ to decrease at a limited rate. For instance, Theorem  \ref{thm:subgroup} continues to holds if $c_5 \rightarrow 0$ and $c_5T^{1/2-c_f} \rightarrow \infty$. Finally, we comment that, the imposed subgrouping structure encourages grouping of similar response processes, which helps further integrate the information across different individual processes.

\section{Simulations}
\label{sec:simulations}

\subsection{Model with low-rank and sparsity structures}
\label{sec:sim1}

We study the finite-sample performance of the proposed method under different predictor processes, link functions $\phi$, point process dimensions $m, p$ and time length $T$. We first consider a model with the low-rank and sparsity structure in this section, then a model with an additional subgroup structure in the next section. 

\begin{figure}[t]
\begin{center}
\begin{tabular}{ccc}
\includegraphics[width=0.3\linewidth]{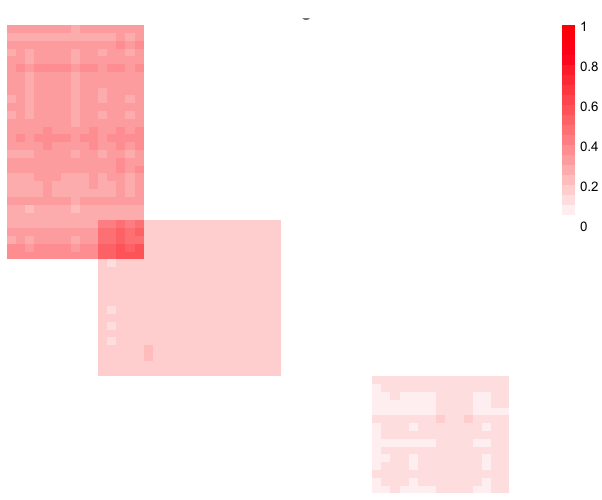} & &
\includegraphics[width=0.3\linewidth]{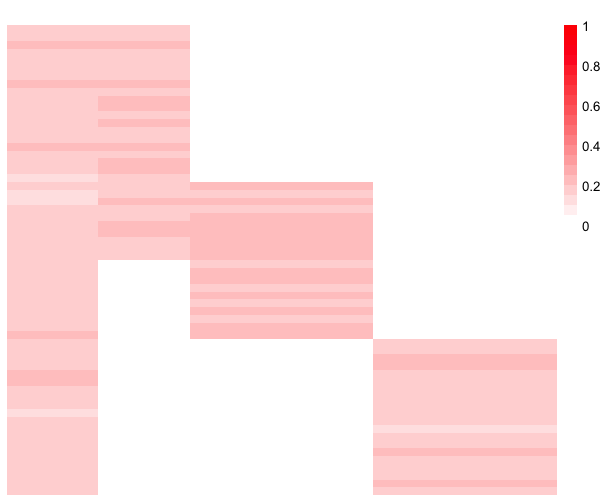} \\
(a) True model coefficients  in Section \ref{sec:sim1} & &
(b) True model coefficients in Section \ref{sec:sim2} 
\end{tabular}
\end{center}
\vspace{-0.1in}
\caption{The slice of the true transferring coefficient tensor $\bcB$, where each entry is $\|\bbeta_{ij}^{\cdot}\|_2$.}
\label{fig:trueB}
\end{figure}

We generate the data following model (\ref{eq:int_b}). Specifically, we first generate the $p$-dimensional predictor point process $\bXt$. We consider two predictor processes, a homogeneous Poisson process with the marginal intensity $\Lambda^x_j$, and a Hawkes process with the transferring function $\omega_{jj'}(t)=a_{jj'} e^{-\beta t}$ and the initial intensity $\Lambda^{(0)}_j$, where $\alpha_{jj'}$ is generated from a uniform distribution on $[0.2, 0.3]$, $\beta=0.7$, and $j, j' = 1, \ldots, p$. We consider two intensity link functions $\phi$, a linear link and a logit link. For the linear link, we set the marginal intensity $\Lambda^x_j=0.5$ for the Poisson predictor process, and set the initial intensity $\Lambda^{(0)}_j=0.3$ for the Hawkes predictor process, $j=1,\ldots,p$. For the logit link, we set $\Lambda^x_j=0.2$ for the Poisson process, and set $\Lambda^{(0)}_j=0.15$ for the Hawkes process, $j=1,\ldots,p$. This way, the Poisson and Hawkes predictor processes are generated with similar levels of overall intensities. Next, we employ a mixture of three basis functions, $g^{(1)}(t)=\exp(-5t)$, $g^{(2)}(t)=0.2 \, \bm{1}(t \le 0.1)$, and $g^{(3)}(t)=0.05 \, \bm{1}(t \le 1)$. The first basis function is an exponential decaying kernel that is widely used in point process modeling. The other two basis functions are piecewise indicator functions, and they are used to capture some ``short-term'' effect and `` long-term'' effect, respectively, that are motivated by neuronal spike trains analysis. Next, we generate the transferring coefficient tensor $\bcB$ with a rank-3 structure, $\bcB=\sum_{r=1}^3 \nu_r \bb^y_r \circ \bb^x_r \circ \bb^c_r$. For the linear link, we set $\nu=(0.3,0.2,0.3)\trans$, 
\vspace{-0.05in}
\begin{center}
\begin{tabular}{ll}
$\bb^y_1=\left( (\bm{\eta}^y_1)_{m/2}\trans, \bm{0}_{m/2}\trans \right)\trans$,  &  $\bb^x_1=\left( (\bm{\eta}^x_1)_{p/3}\trans, \bm{0}_{3p/4}\trans \right)\trans$, \\
$\bb^y_2=\left( \bm{0}_{5m/12}\trans, (\bm{\eta}^y_2)_{m/3}\trans,\bm{0}_{m/4}\trans \right)\trans$,  &  $\bb^x_2=\left( \bm{0}_{p/6}\trans,(\bm{\eta}^x_2)_{p/3}\trans, \bm{0}_{p/2}\trans \right)\trans$, \\
$\bb^y_3=\left( \bm{0}_{3m/4}\trans, (\bm{\eta}^y_3)_{m/4}\trans \right)\trans$,  &  $\bb^x_3=\left( \bm{0}_{2p/3}\trans, (\bm{\eta}^x_3)_{p/4}\trans, \bm{0}_{p/12}\trans \right)\trans$,
\end{tabular}
\end{center}
and $\bb^c_r$, $\bm{\eta}^y_r$ and $\bm{\eta}^x_r$, $r=1,2,3$, are all generated from a normal distribution with mean one and covariance the identity matrix. Figure \ref{fig:trueB}(a) shows the true association structure based on the generated coefficient $\bcB$. For the logit link, we set $\nu=(0.2,0.1,0.2)\trans$, and generate $\bb^y_r, \bb^x_r, \bb^c_r$ in the same way as for the linear link, except that we add a negative sign to each element of $\bcB$ with probability 0.5. Finally, we set the background intensity $\bmu=\bm{0.01}_m$, then generate the $m$-dimensional response point process $\bYt$ following model (\ref{eq:int_b}). Given the intensity function, each individual response process is simulated following the thinning strategy \citep{ogata1988}. We set the dimension of the response and predictor process $m = p = \{60, 120\}$, and the observed length $T = \{800, 2000\}$. For a homogeneous process, $T$ plays the role of sample size, since it is proportional to the expected number of events. For an inhomogeneous process, this is not necessarily true, and the expected number of observed events could vary across different marginal processes.

We compare our method with two alternative solutions. The first is a standard baseline model that simply fits the conditional intensity functions in \eqref{eq:int_f} without specifying any additional structure. The second adds a group $\ell_1$ penalty on the transferring coefficients to the baseline model for sparsity-pursuit, which is analogous to \cite{hansen2015}.  We evaluate the estimation accuracy by the root mean square error (RMSE) of the estimated coefficient tensor $\bcB$.

\begin{table}[t]
\centering
\caption{Estimation accuracy of the transferring coefficient tensor $\bcB$ for the model in Section \ref{sec:sim1}. Three methods are compared: the regular point process regression model (PP-Reg), the sparse point process regression with a group $\ell_1$ penalty (SpPP-Reg), and our proposed method.  Reported are the average RMSE based on 50 replications, with the standard errors in the parenthesis.}
\label{tb:sim1}
\begin{tabular}{llcc|ccc}
\hline
Link & Predictor & $m=p$ & $T$ & PP-Reg & SpPP-Reg & Our method \\ \hline
\multirow{8}{*}{Linear}&\multirow{4}{*}{Poisson}&\multirow{2}{*}{60}  & 800 & 0.281 (0.019) &  0.234 (0.010) &  0.147 (0.011)   \\
  &  &  & 2000 & 0.168 (0.010) & 0.149 (0.007)  & 0.094 (0.006) \\ \cline{3-7}
  &  &  \multirow{2}{*}{120} & 800 & 0.319 (0.025) & 0.263 (0.021) & 0.117 (0.015)  \\
  &  &  & 2000 & 0.189 (0.011) & 0.169 (0.009) & 0.066 (0.009)  \\ \cline{2-7}
  & \multirow{4}{*}{Hawkes} & \multirow{2}{*}{60} & 800 & 0.307 (0.045) & 0.279 (0.028) & 0.185 (0.025) \\
  &  &  & 2000 & 0.226 (0.026)  & 0.197 (0.021) & 0.125 (0.018) \\ \cline{3-7}
  &  &  \multirow{2}{*}{120}	 & 800 & 0.337 (0.034) & 0.289 (0.024) & 0.129  (0.016) \\
  &  &  & 2000 & 0.245 (0.015)  &  0.205 (0.010)  & 0.079 (0.010) \\ \hline                                                                                                      
\multirow{8}{*}{Logit}&\multirow{4}{*}{Poisson} &\multirow{2}{*}{60}  & 800 & 0.548 (0.026) & 0.247 (0.015) & 0.152 (0.012) \\
  &  &  & 2000 & 0.518 (0.015)  & 0.202 (0.009) & 0.121 (0.009) \\ \cline{3-7} 
  &  & \multirow{2}{*}{120} 	& 800 & 0.844  (0.065)  &  0.264 (0.025)  & 0.134  (0.015) \\
  &  & & 2000 & 0.645 (0.017) & 0.196 (0.005)  & 0.101 (0.003)  \\ \cline{2-7}
  & \multirow{4}{*}{Hawkes} & \multirow{2}{*}{60}  & 800 & 0.648 (0.045)  & 0.276 (0.028) & 0.158 (0.025) \\
  & &  & 2000 &0.583 (0.035)  & 0.192 (0.018) & 0.124 (0.012)  \\ \cline{3-7}
  & & \multirow{2}{*}{120} & 800 & 0.983 (0.048) &  0.306 (0.026) & 0.149 (0.014)  \\
  & &	 & 2000 & 0.725 (0.026) &  0.211 (0.017)  & 0.103  (0.016)  \\
\hline
\end{tabular}
\end{table}

\begin{figure}[t!]
\begin{center}
\begin{tabular}{ccc}
\includegraphics[width=0.3\linewidth]{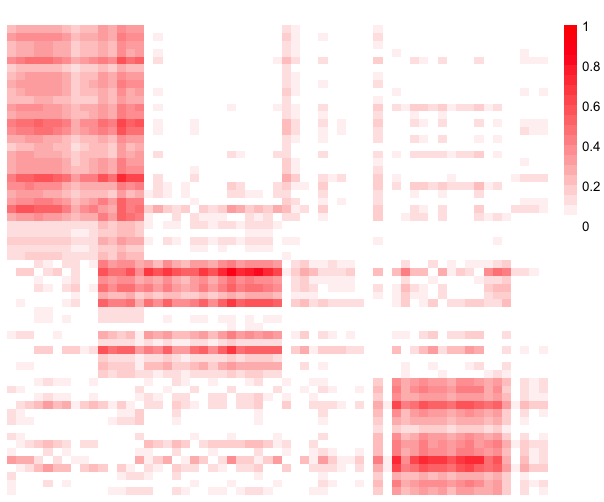} &
\includegraphics[width=0.3\linewidth]{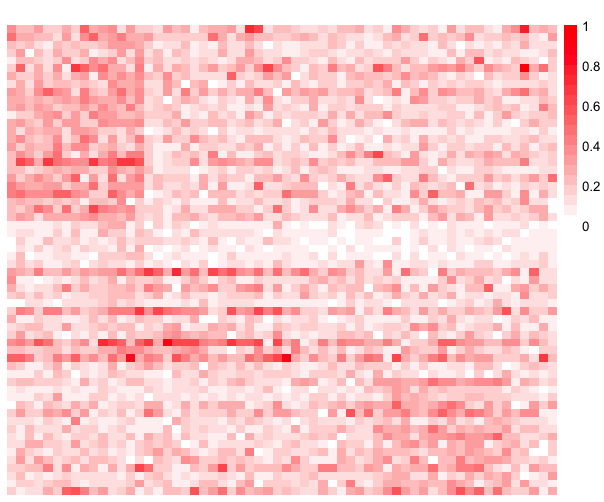} &
\includegraphics[width=0.3\linewidth]{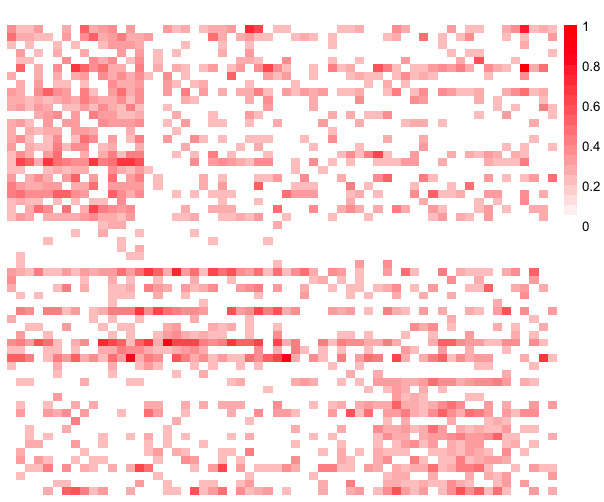} \\
{\small Our method: $m=60,T=800$} & 
{\small PP-Reg: $m=60,T=800$} &
{\small SpPP-Reg: $m=60,T=800$} \\
\includegraphics[width=0.3\linewidth]{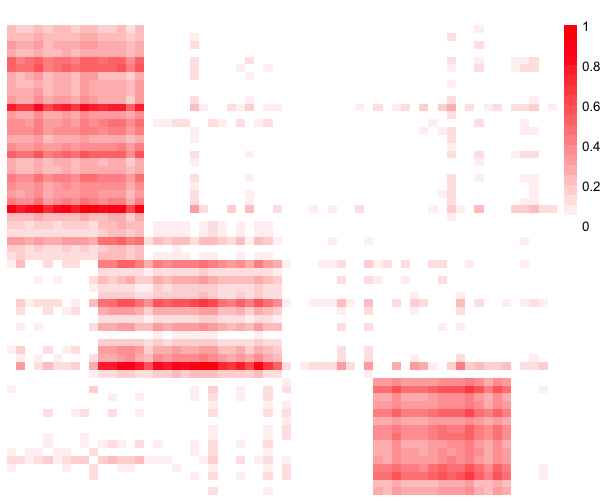} &
\includegraphics[width=0.3\linewidth]{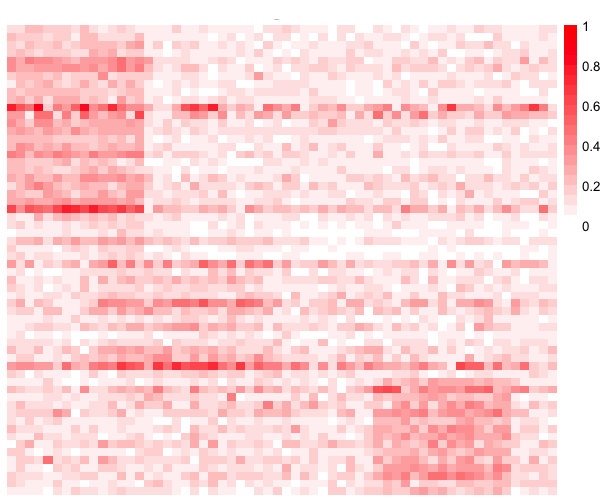} & 
\includegraphics[width=0.3\linewidth]{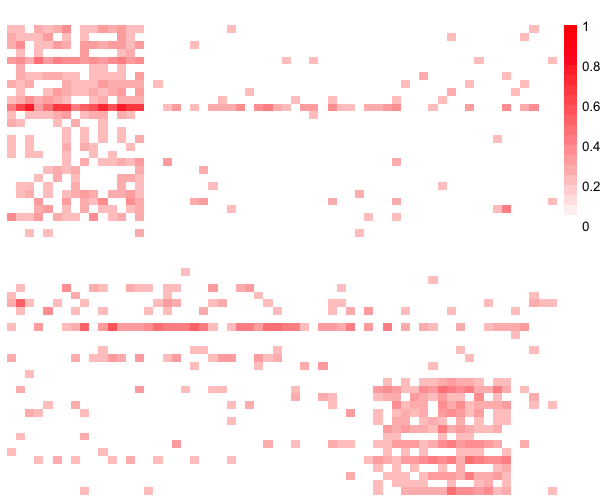} \\
{\small Our method: $m=60,T=2000$} & 
{\small PP-Reg: $m=60,T=2000$} &
{\small SpPP-Reg: $m=60,T=2000$} \\
\includegraphics[width=0.3\linewidth]{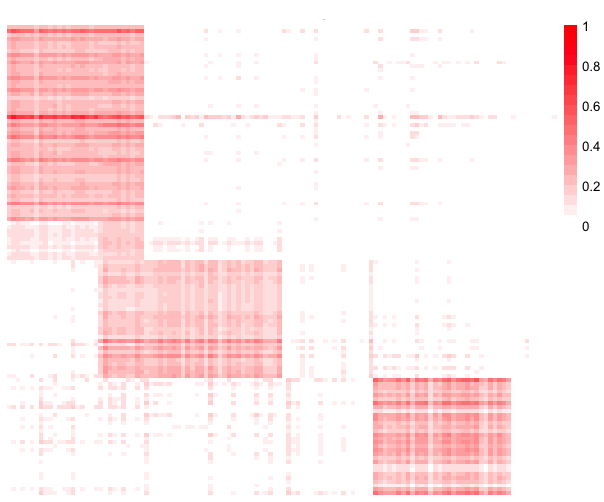} & 
\includegraphics[width=0.3\linewidth]{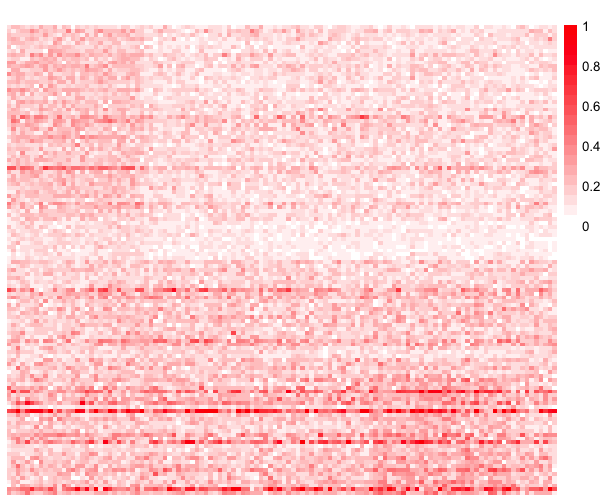} & 
\includegraphics[width=0.3\linewidth]{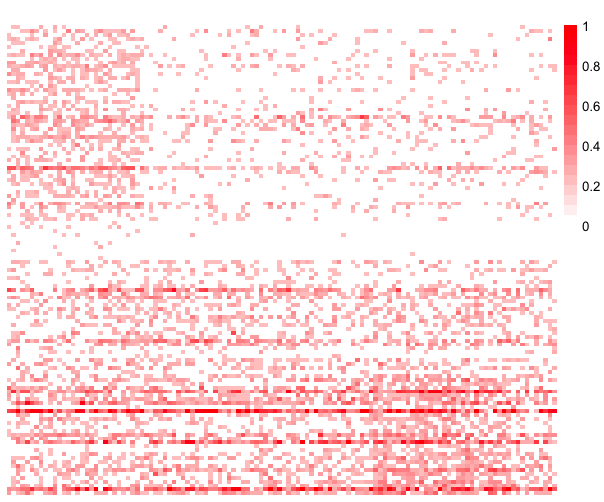} \\
{\small Our method: $m=120,T=2000$} & 
{\small PP-Reg: $m=120,T=2000$} &
{\small SpPP-Reg: $m=120,T=2000$} 
\end{tabular}
\end{center}
\caption{Recovered transferring coefficient tensor $\bcB$ for the model in Section \ref{sec:sim1}. Three methods are compared: the regular point process regression model (PP-Reg), the sparse point process regression with a group $\ell_1$ penalty (SpPP-Reg), and our proposed method.}
\label{fig:sim1}
\end{figure}

Table \ref{tb:sim1} summarizes the results based on 50 data replications. It is seen that our proposed method consistently outperforms the two alternative solutions, as it achieves the smallest RMSE across all settings. As the point process length $T$ increases, all methods improve in estimation accuracy. On the other hand, as the numbers of point processes $m$ and $p$ increase, our method continues to improve, whereas the two alternative solutions suffer. This is largely due to that our method jointly model all the point processes together, instead of one at a time. Figure \ref{fig:sim1} shows the recovered transferring structure based on the estimated $\bcB$ with a linear link and a Poisson predictor process. It is seen again that our method is capable of recovering the transferring structure successfully, while the other two methods cannot.

\subsection{Model with additional subgrouping structure}
\label{sec:sim2}

We next consider a model with an additional subgrouping structure. For simplicity, we focus on the linear link $\phi$ and the Poisson predictor process. The results are similar for other combinations of link function and predictor process. We adopt the same simulation setup as in Section \ref{sec:sim1}, except that we generate the transferring coefficient tensor $\bcB$ in a different way. Specifically, we consider a rank-4 structure $\bcB=\sum_{r=1}^4 \nu_r \bb^y_r \circ \bb^x_r \circ \bb^c_r$. We set $\bnu=(0.2, 0.2, 0.2, 0.2)\trans$,  
\begin{center}
\begin{tabular}{ll}
$\bb^y_1=\left( (\eta^y_1 \bm{1})_{p/6}\trans, \bm{0}_{5p/6}\trans \right)\trans$,  &  $\bb^x_1=(\bm{\eta}^x_1)_{m}$, \\
$\bb^y_2=\left( \bm{0}_{p/6}\trans,(\eta^y_2 \bm{1})_{p/6}\trans, \bm{0}_{2p/3}\trans \right)\trans$,  &  $\bb^x_2=\left( (\bm{\eta}^x_2)_{m/2}\trans, \bm{0}_{m/2}\trans \right)\trans $, \\
$\bb^y_3=\left( \bm{0}_{p/3}\trans, (\eta^y_3 \bm{1})_{p/3}\trans, \bm{0}_{p/3}\trans \right)\trans$,  &  $\bb^x_3=\left( \bm{0}_{m/3}\trans, (\bm{\eta}^x_3)_{m/3}\trans, \bm{0}_{m/3}\trans \right)\trans $, \\
$\bb^y_4=\left( \bm{0}_{2p/3}\trans, (\eta^y_4 \bm{1})_{p/3}\trans \right)\trans$,  &  $\bb^x_4=\left( \bm{0}_{2m/3}\trans, (\bm{\eta}^x_4)_{m/3}\trans \right)\trans$, 
\end{tabular}
\end{center}
$\bb^c_r$, $\bm{\eta}^x_r$ are all generated from a normal distribution with mean one and the identity covariance, and $\eta^y_r$, $r=1,2,3,4$, are generated from a univariate normal distribution with mean one and variance $0.1$. Note that, unlike the coefficient tensor in Section \ref{sec:sim1}, here the entries are repeated in $\bb^y_r$, which in turn induces the subgrouping structure. This structure can also be seen in Figure \ref{fig:trueB}(b), which shows a slice of one generated coefficient tensor $\bcB$. We set the dimension of the response and predictor process $m = p = \{60, 120\}$, and the observed length $T=\{1200, 2400\}$.  

\begin{table}[t!]
\centering
\caption{Estimation and clustering accuracy of the transferring coefficient tensor $\bcB$ for the model in Section \ref{sec:sim2}. Three methods are compared: the regular point process regression model (PP-Reg), the sparse point process regression with a group $\ell_1$ penalty (SpPP-Reg), and our proposed method. Reported are the average RMSE based on 50 replications, with the standard errors in the parenthesis. Rand index for our method is also reported.}
\label{tb:sim2}
\begin{tabular}{cc|cccc}
\hline
$m=p$   & $T$ &  PP-Reg & SpPP-Reg  &  Our method & \{Rand Index\}  \\ \hline
\multirow{2}{*}{60}   & 1200 & 0.179 (0.026) &  0.162 (0.013) &  0.088 (0.010) &\{0.854 (0.063)\} \\
                                & 2400 & 0.142 (0.012) &  0.136 (0.008) &  0.066 (0.005) &\{0.909 (0.054)\} \\\hline
\multirow{2}{*}{120} & 1200 & 0.219 (0.021) &  0.179 (0.018) & 0.073 (0.014) &\{0.878 (0.099)\}  \\
                                & 2400 & 0.169 (0.008) &  0.147 (0.005) & 0.052 (0.003) &\{0.912 (0.071)\} \\
\hline
\end{tabular}
\end{table}

\begin{figure}[t!]
\begin{center}
\begin{tabular}{ccc}
\includegraphics[width=0.3\linewidth]{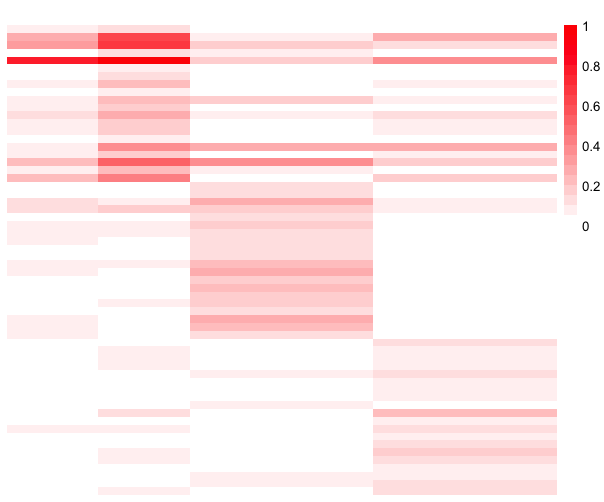} &
\includegraphics[width=0.3\linewidth]{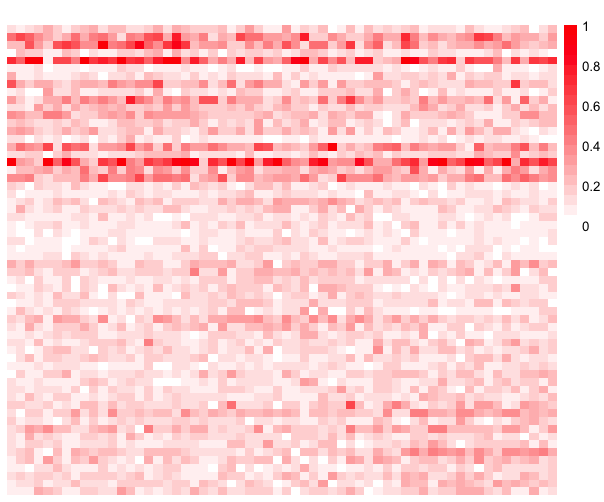} &
\includegraphics[width=0.3\linewidth]{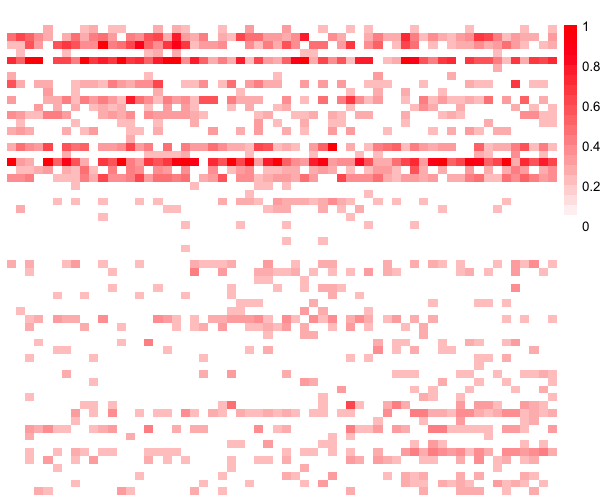} \\
{\small Our method: $m=60,T=1200$} & 
{\small PP-Reg: $m=60,T=1200$} &
{\small SpPP-Reg: $m=60,T=1200$} \\
\includegraphics[width=0.3\linewidth]{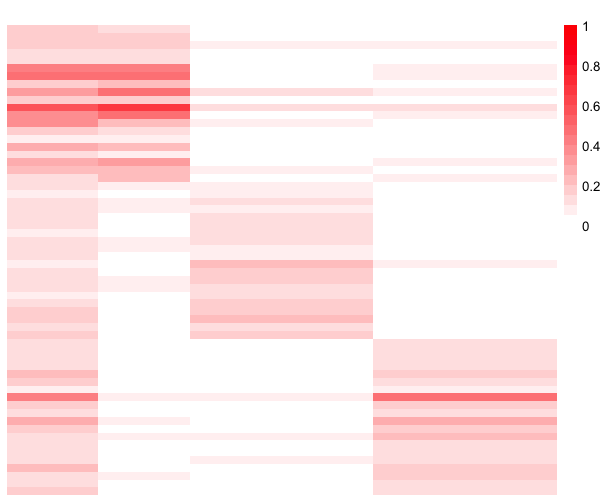} &
\includegraphics[width=0.3\linewidth]{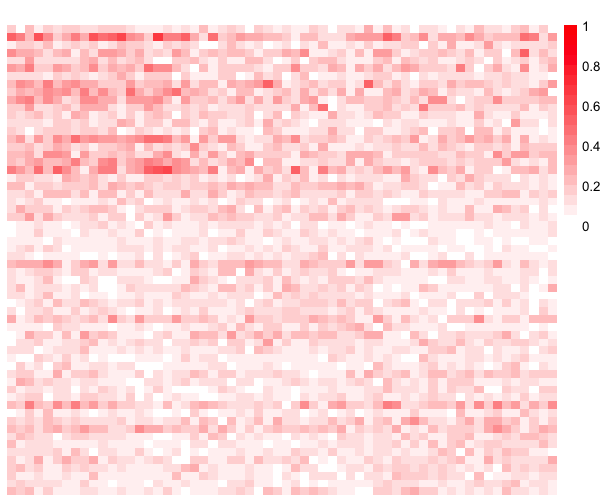} & 
\includegraphics[width=0.3\linewidth]{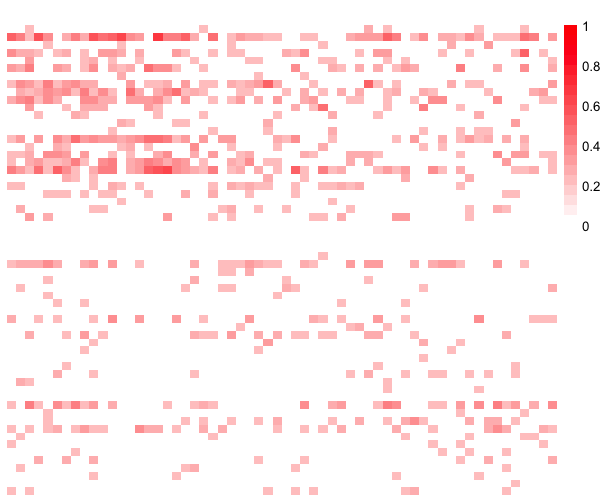} \\
{\small Our method: $m=60,T=2400$} & 
{\small PP-Reg: $m=60,T=2400$} &
{\small SpPP-Reg: $m=60,T=2400$} \\
\includegraphics[width=0.3\linewidth]{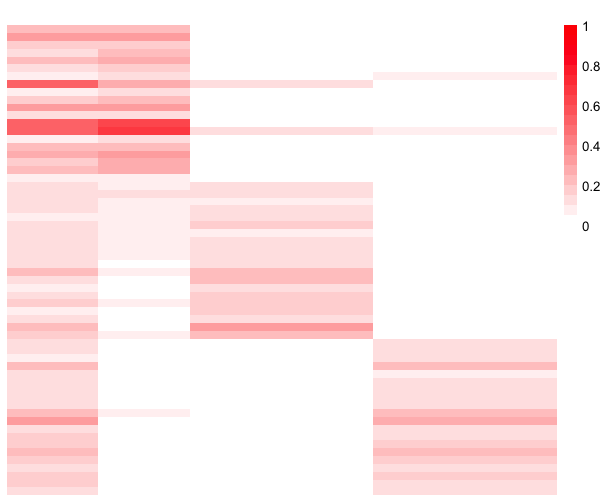} & 
\includegraphics[width=0.3\linewidth]{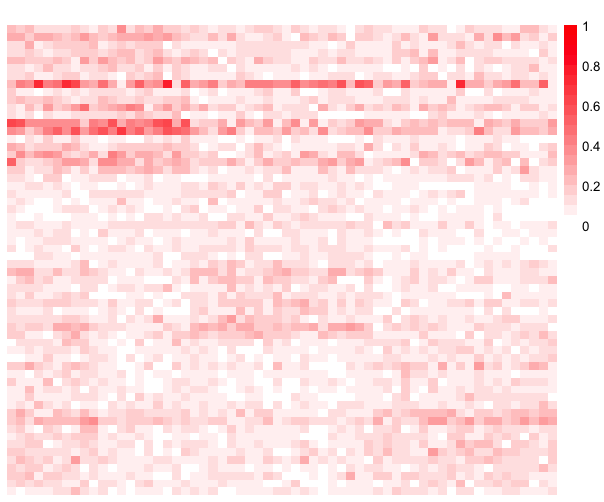} & 
\includegraphics[width=0.3\linewidth]{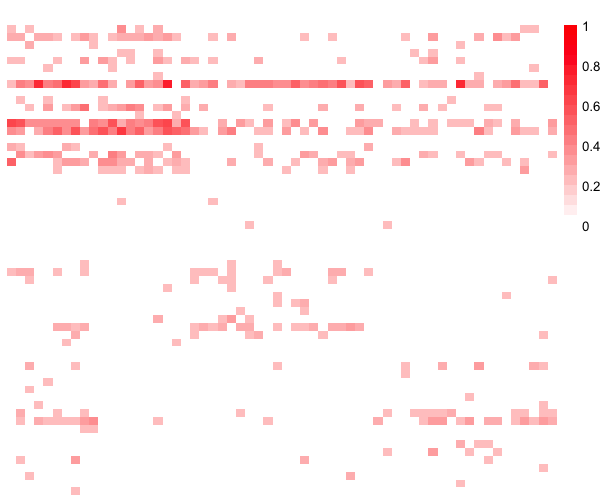} \\
{\small Our method: $m=120,T=2400$} & 
{\small PP-Reg: $m=120,T=2400$} &
{\small SpPP-Reg: $m=120,T=2400$} 
\end{tabular}
\end{center}
\caption{Recovered transferring coefficient tensor $\bcB$ for the model in Section \ref{sec:sim2}. Three methods are compared: the regular point process regression model (PP-Reg), the sparse point process regression with a group $\ell_1$ penalty (SpPP-Reg), and our proposed method.}
\label{fig:sim2}
\end{figure}

Table \ref{tb:sim2} summarizes RMSE based on 50 data replications, and Figure \ref{fig:sim2} shows recovered transferring structure. It is again seen that our proposed method consistently outperforms the two alternative solutions in terms of estimation accuracy. Moreover, Table \ref{tb:sim2} includes the rand index statistic for our proposed method, which evaluates the clustering performance. It is seen that our method achieves a high index value in all settings. 

As for the computation time, for the simulation example in Section \ref{sec:sim1} with a linear link, a Poisson predictor process, $m=p=60$ and $T=2000$, the average computing time was about 1.1 minutes. For the example in Section \ref{sec:sim2} with $m=p=60$ and $T=2400$, the average computing time was about 1.9 minutes. All computations were done on a personal laptop with Intel(R) Core(TM) i7-8565U CPU@1.8GHz.

\section{Cross-area Neuronal Spike Trains Analysis}
\label{sec:realdata}

Ensemble neural activity analysis is of central importance in system neuroscience, which aims to understand sensory coding and associations with motor output and cognitive functions \citep{brown2004, kim2011granger}. Some goals of common interest include the study of single-neuron activity with dependence on its own history, and the study of cross-neuron correlations based on spike trains similarities within the same area.  Beyond those goals, it is also of key interest to understand the communication patterns in information transmission between neurons in different brain areas through neuronal spiking activities \citep{saalmann2012}. A group of neurons could be identified within a brain area based on their similar exciting or inhibitory effects onto another group of neurons in a different brain area.  This hypothesis has been suggested by several scientific studies. For instance, \citet{liang2013}  found that there might be discrete locations within the visual cortex area that respond to specific cross-modal inputs such as auditory or tactile. That is, the neurons in the V1 area are expected to be clustered in that they share similar cross-cortex-area association patterns, which needs  to be inferred from the associations between the observed spike trains activities. In addition, the signal transmission takes time from one area to another, suggesting that the cross-area neuronal connection may account for a time-dependent convolutional effect rather than a simple co-firing.  In recent years, benefitting from the rapid development of imaging techniques such as the calcium imaging, we are now able to monitor a large number of neurons simultaneously with a single-neuron resolution in a short time period, which produces high-dimensional point process type data of neuronal spike trains. 

\begin{figure}[t]
\begin{center}
\begin{tabular}{ccc}
\includegraphics[width=0.4\linewidth]{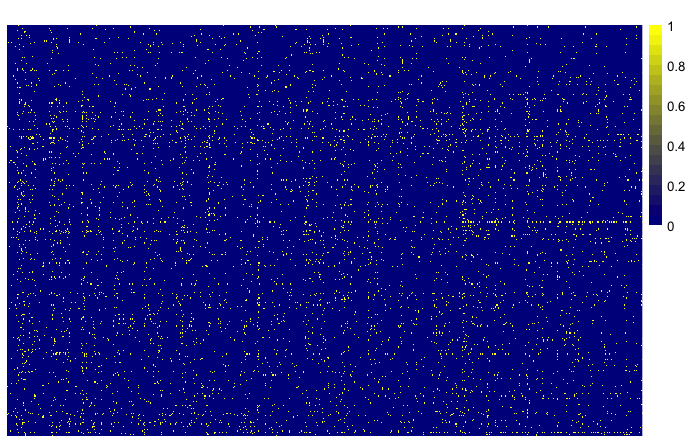} & & \includegraphics[width=0.4\linewidth]{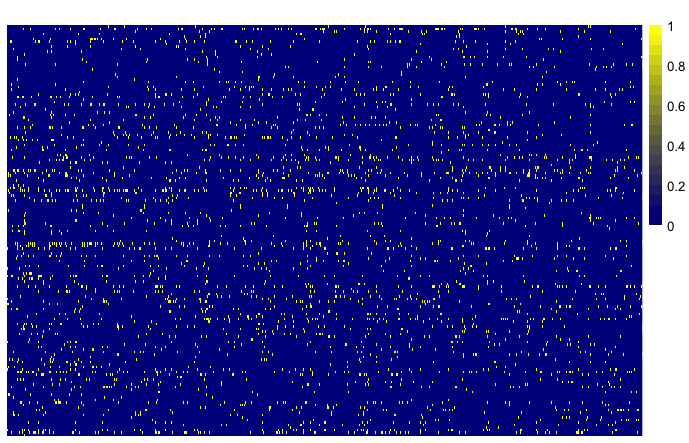} \\ 
\includegraphics[width=0.4\linewidth]{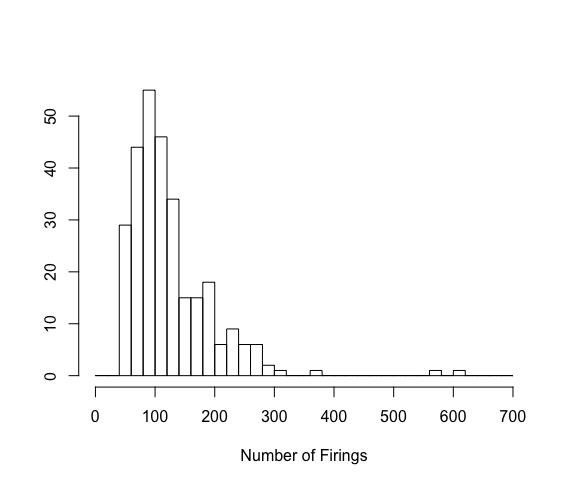} & & \includegraphics[width=0.4\linewidth]{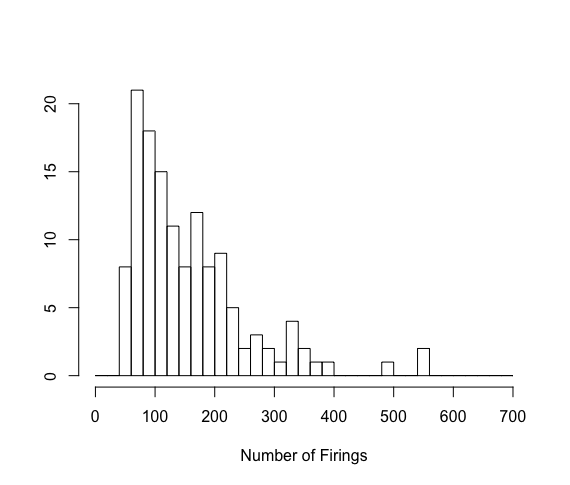} \\
(a) V1 & & (b) A1
\end{tabular}
\end{center}
\vspace{-0.1in}
\caption{The neuron firings in the V1 and A1 areas. The heatmaps (upper panels) show the neuron-wise firings over time. The histograms (lower panels) summarize the number of firings for each neuron. }
\label{fig:A1V1}
\end{figure}

In our study, we simultaneously measure the neuronal spike trains activities of 139 neurons and 283 neurons from two sensory cortical areas, A1 and V1,  in a rat brain, respectively. We collect the data over 192 seconds under a stable stimulus. With 50 millisecond as a unit of time, we obtain the length of time interval of $[0,3840]$. Figure \ref{fig:A1V1} shows the recorded neuron firing events over time, and the histogram summary of the numbers of observed firings for individual neurons in each of these two areas. It is seen that  most neurons have their numbers of observed firing events under 200, whereas a subset of neurons have the numbers below 100.

Since a primary goal is to understand the information transmission from the A1 area to the V1 area, we fit the data using our proposed multivariate temporal point process regression, by treating the neuronal spike trains in V1 as the response point process, and the neuronal spike trains in A1 as the predictor process. Since the observed firing events are sparse, we choose a logit link function. We select three basis functions, similarly as in our simulation studies: $g^{(1)}(t)=\exp(-t)$, $g^{(2)}(t)=0.2 \, \bm{1}(t \le 1)$, and $g^{(3)}(t)=0.05 \, \bm{1}\{t\le 5\}$, with the time intervals in the indicator functions selected based on the existing scientific findings that the communication process between ensemble neurons across areas mostly happens within tens of milliseconds \citep{luo2016}. In addition to our proposed model, we also fit the marginal model that takes one response process at a time. Since some neurons have very limited number of firing events, the corresponding model fittings may not converge. Actually, for our data, we have found that about one third of the individual response process fittings cannot converge. To handle this convergence issue, we add an $\ell_2$ regularization to this marginal approach, though we still refer to it as a marginal method. Moreover, we fit the marginal model with a group $\ell_1$ regularization, similarly as in our simulations. 

To evaluate the model, we split the point processes into a training set, i.e., the spike trains data in the time interval $[0,2000)$, and a testing set, i.e., the data in the time interval  $(2000,3800]$. We report two evaluation criteria. The first criterion is the area under the ROC curve (AUC) based on a binary prediction \citep{luo2016}. That is, we bin the continuous point process into a sequence of binary values based on a unit of time of 50 milliseconds, with one meaning that there is a firing event in this time bin, and zero otherwise. We then produce a sequence of binary predictions based on the predicted intensity function for the testing data. The second criterion is the deviance $\|\hat{\bcB}_{training} - \hat{\bcB}_{testing}\|$. That is, we obtain the estimated coefficient tensor $\bcB$ from the training data and testing data, respectively, and evaluate the difference between the two in the Frobenius norm. Intuitively, if the firing patterns have been consistent, then this deviance measure should be small. Table \ref{tb:real} reports the results. It is seen that our proposed method achieves the highest AUC value and the lowest deviance value, suggesting a competitive performance of the proposed method compared to the two alternative solutions. We also identify five subgroups of neurons with our method, which requires future  scientific validation, as we do not have relevant subgroup information for this dataset. 

\begin{table}[t!]
\centering
\caption{Evaluation of the model fitting for the cross-area neuronal spike trains analysis. Three methods are compared: the regular point process regression model (PP-Reg), the sparse point process regression with a group $\ell_1$ penalty (SpPP-Reg), and our proposed method.}
\label{tb:real}
\begin{tabular}{l|ccc} \hline
 & PP-Reg & SpPP-Reg & Our method \\ \hline
Deviance &  0.388  &  0.256  &  0.185   \\
AUC        &  0.537  &  0.579  &  0.682   \\ \hline
\end{tabular}
\end{table}

\baselineskip=17pt 
\bibliographystyle{apalike}
\bibliography{ref-ppreg}

\begin{thebibliography}{}

\bibitem[Bacry et~al., 2020]{Bacry2020}
Bacry, E., Bompaire, M., Ga{\"\i}ffas, S., and Muzy, J.-F. (2020).
\newblock Sparse and low-rank multivariate hawkes processes.
\newblock {\em Journal of Machine Learning Research}, 21:1--32.

\bibitem[Bacry and Muzy, 2016]{bacry2016}
Bacry, E. and Muzy, J.-F. (2016).
\newblock First-and second-order statistics characterization of hawkes
  processes and non-parametric estimation.
\newblock {\em IEEE Transactions on Information Theory}, 62(4):2184--2202.

\bibitem[Bi et~al., 2018]{bi2018}
Bi, X., Qu, A., Shen, X., et~al. (2018).
\newblock Multilayer tensor factorization with applications to recommender
  systems.
\newblock {\em The Annals of Statistics}, 46(6B):3308--3333.

\bibitem[Boyd et~al., 2011]{boyd2011}
Boyd, S., Parikh, N., Chu, E., Peleato, B., Eckstein, J., et~al. (2011).
\newblock Distributed optimization and statistical learning via the alternating
  direction method of multipliers.
\newblock {\em Foundations and Trends{\textregistered} in Machine learning},
  3(1):1--122.

\bibitem[Br{\'e}maud and Massouli{\'e}, 1996]{bremaud1996}
Br{\'e}maud, P. and Massouli{\'e}, L. (1996).
\newblock Stability of nonlinear hawkes processes.
\newblock {\em The Annals of Probability}, pages 1563--1588.

\bibitem[Brown et~al., 2004]{brown2004}
Brown, E.~N., Kass, R.~E., and Mitra, P.~P. (2004).
\newblock Multiple neural spike train data analysis: state-of-the-art and
  future challenges.
\newblock {\em Nature neuroscience}, 7(5):456.

\bibitem[Chen et~al., 2012]{chen2012}
Chen, B., He, S., Li, Z., and Zhang, S. (2012).
\newblock Maximum block improvement and polynomial optimization.
\newblock {\em SIAM Journal on Optimization}, 22(1):87--107.

\bibitem[Chen et~al., 2019a]{ChenYuan2019}
Chen, H., Raskutti, G., and Yuan, M. (2019a).
\newblock Non-convex projected gradient descent for generalized low-rank tensor
  regression.
\newblock {\em Journal of Machine Learning Research}, 20(5):1--37.

\bibitem[Chen et~al., 2019b]{chen2019}
Chen, S., Shojaie, A., Shea-Brown, E., and Witten, D. (2019b).
\newblock The multivariate hawkes process in high dimensions: Beyond mutual
  excitation.
\newblock {\em arXiv preprint arXiv:1707.04928}.

\bibitem[Daley and Vere-Jones, 2007]{daley2007}
Daley, D.~J. and Vere-Jones, D. (2007).
\newblock {\em An introduction to the theory of point processes: volume II:
  general theory and structure}.
\newblock Springer Science \& Business Media.

\bibitem[Deng et~al., 2017]{GuanZhang2017}
Deng, C., Guan, Y., Waagepetersen, R.~P., and Zhang, J. (2017).
\newblock Second-order quasi-likelihood for spatial point processes.
\newblock {\em Biometrics}, 73(4):1311--1320.

\bibitem[Diggle et~al., 2010]{diggle2010risk}
Diggle, P.~J., Guan, Y., Hart, A.~C., Paize, F., and Stanton, M. (2010).
\newblock Estimating individual-level risk in spatial epidemiology using
  spatially aggregated information on the population at risk.
\newblock {\em Journal of the American Statistical Association},
  105(492):1394--1402.

\bibitem[Guan, 2008]{guan2008consistent}
Guan, Y. (2008).
\newblock On consistent nonparametric intensity estimation for inhomogeneous
  spatial point processes.
\newblock {\em Journal of the American Statistical Association},
  103(483):1238--1247.

\bibitem[Guan, 2011]{guan2011second}
Guan, Y. (2011).
\newblock Second-order analysis of semiparametric recurrent event processes.
\newblock {\em Biometrics}, 67(3):730--739.

\bibitem[Guan et~al., 2015]{guan2015quasi}
Guan, Y., Jalilian, A., and Waagepetersen, R. (2015).
\newblock Quasi-likelihood for spatial point processes.
\newblock {\em Journal of the Royal Statistical Society: Series B (Statistical
  Methodology)}, 77(3):677--697.

\bibitem[Hansen et~al., 2015]{hansen2015}
Hansen, N.~R., Reynaud-Bouret, P., Rivoirard, V., et~al. (2015).
\newblock Lasso and probabilistic inequalities for multivariate point
  processes.
\newblock {\em Bernoulli}, 21(1):83--143.

\bibitem[Hawkes, 1971]{hawkes1971}
Hawkes, A.~G. (1971).
\newblock Spectra of some self-exciting and mutually exciting point processes.
\newblock {\em Biometrika}, 58(1):83--90.

\bibitem[Ji et~al., 2016]{ji2016}
Ji, N., Freeman, J., and Smith, S.~L. (2016).
\newblock Technologies for imaging neural activity in large volumes.
\newblock {\em Nature neuroscience}, 19(9):1154.

\bibitem[Kang et~al., 2011]{Kang2011}
Kang, J., Johnson, T.~D., Nichols, T.~E., and Wager, T.~D. (2011).
\newblock Meta analysis of functional neuroimaging data via bayesian spatial
  point processes.
\newblock {\em Journal of the American Statistical Association},
  106(493):124--134.

\bibitem[Kang et~al., 2014]{Kang2014}
Kang, J., Nichols, T., Wager, T., and Johnson, T. (2014).
\newblock A bayesian hierarchical spatial point process model for multi-type
  neuroimaging meta-analysis.
\newblock {\em The Annals of Applied Statistics}, 8:1800--1824.

\bibitem[Kim et~al., 2011]{kim2011granger}
Kim, S., Putrino, D., Ghosh, S., and Brown, E.~N. (2011).
\newblock A granger causality measure for point process models of ensemble
  neural spiking activity.
\newblock {\em PLoS computational biology}, 7(3):e1001110.

\bibitem[Kolda and Bader, 2009]{kolda2009}
Kolda, T.~G. and Bader, B.~W. (2009).
\newblock Tensor decompositions and applications.
\newblock {\em SIAM review}, 51(3):455--500.

\bibitem[Liang et~al., 2013]{liang2013}
Liang, M., Mouraux, A., Hu, L., and Iannetti, G. (2013).
\newblock Primary sensory cortices contain distinguishable spatial patterns of
  activity for each sense.
\newblock {\em Nature communications}, 4:1979.

\bibitem[Luo et~al., 2016]{luo2016}
Luo, X., Gee, S., Sohal, V., and Small, D. (2016).
\newblock A point-process response model for spike trains from single neurons
  in neural circuits under optogenetic stimulation.
\newblock {\em Statistics in medicine}, 35(3):455--474.

\bibitem[Ma and Huang, 2017]{ma2017subgroup}
Ma, S. and Huang, J. (2017).
\newblock A concave pairwise fusion approach to subgroup analysis.
\newblock {\em Journal of the American Statistical Association},
  112(517):410--423.

\bibitem[Ogata, 1988]{ogata1988}
Ogata, Y. (1988).
\newblock Statistical models for earthquake occurrences and residual analysis
  for point processes.
\newblock {\em Journal of the American Statistical association}, 83(401):9--27.

\bibitem[Ogata et~al., 1978]{ogata1978}
Ogata, Y. et~al. (1978).
\newblock The asymptotic behaviour of maximum likelihood estimators for
  stationary point processes.
\newblock {\em Annals of the Institute of Statistical Mathematics},
  30(1):243--261.

\bibitem[Okun et~al., 2015]{okun2015}
Okun, M., Steinmetz, N.~A., Cossell, L., Iacaruso, M.~F., Ko, H., Barth{\'o},
  P., Moore, T., Hofer, S.~B., Mrsic-Flogel, T.~D., Carandini, M., et~al.
  (2015).
\newblock Diverse coupling of neurons to populations in sensory cortex.
\newblock {\em Nature}, 521(7553):511.

\bibitem[Perry and Wolfe, 2013]{perry2013}
Perry, P.~O. and Wolfe, P.~J. (2013).
\newblock Point process modelling for directed interaction networks.
\newblock {\em Journal of the Royal Statistical Society: Series B (Statistical
  Methodology)}, 75(5):821--849.

\bibitem[Rathbun and Cressie, 1994]{rathbun1994}
Rathbun, S.~L. and Cressie, N. (1994).
\newblock Asymptotic properties of estimators for the parameters of spatial
  inhomogeneous poisson point processes.
\newblock {\em Advances in Applied Probability}, 26(1):122--154.

\bibitem[Saalmann et~al., 2012]{saalmann2012}
Saalmann, Y.~B., Pinsk, M.~A., Wang, L., Li, X., and Kastner, S. (2012).
\newblock The pulvinar regulates information transmission between cortical
  areas based on attention demands.
\newblock {\em Science}, 337(6095):753--756.

\bibitem[Shen, 1998]{shen1998}
Shen, X. (1998).
\newblock On the method of penalization.
\newblock {\em Statistica Sinica}, 8(2):337--357.

\bibitem[Shen and Wong, 1994]{shen1994}
Shen, X. and Wong, W.~H. (1994).
\newblock Convergence rate of sieve estimates.
\newblock {\em The Annals of Statistics}, pages 580--615.

\bibitem[Sidiropoulos and Bro, 2000]{sidiropoulos2000}
Sidiropoulos, N.~D. and Bro, R. (2000).
\newblock On the uniqueness of multilinear decomposition of n-way arrays.
\newblock {\em Journal of Chemometrics}, 14(3):229--239.

\bibitem[Stoyan et~al., 2000]{stoyan2000}
Stoyan, D., Penttinen, A., et~al. (2000).
\newblock Recent applications of point process methods in forestry statistics.
\newblock {\em Statistical Science}, 15(1):61--78.

\bibitem[Sun and Li, 2017]{sun2017store}
Sun, W. and Li, L. (2017).
\newblock Store: Sparse tensor response regression and neuroimaging analysis.
\newblock {\em Journal of Machine Learning Research}, 18:4908--4944.

\bibitem[Tang et~al., 2019]{tang2019}
Tang, X., Bi, X., and Qu, A. (2019).
\newblock Individualized multilayer tensor learning with an application in
  imaging analysis.
\newblock {\em Journal of the American Statistical Association},
  115(530):836--851.

\bibitem[Waagepetersen and Guan, 2009]{waagepetersen2009two}
Waagepetersen, R. and Guan, Y. (2009).
\newblock Two-step estimation for inhomogeneous spatial point processes.
\newblock {\em Journal of the Royal Statistical Society: Series B (Statistical
  Methodology)}, 71(3):685--702.

\bibitem[Wang et~al., 2016]{wang2016}
Wang, Y., Xie, B., Du, N., and Song, L. (2016).
\newblock Isotonic hawkes processes.
\newblock In {\em International conference on machine learning}, pages
  2226--2234.

\bibitem[Yuan and Lin, 2006]{yuan2006}
Yuan, M. and Lin, Y. (2006).
\newblock Model selection and estimation in regression with grouped variables.
\newblock {\em Journal of the Royal Statistical Society: Series B (Statistical
  Methodology)}, 68(1):49--67.

\bibitem[Zhang and Han, 2019]{ZhangA2019}
Zhang, A. and Han, R. (2019).
\newblock Optimal sparse singular value decomposition for high-dimensional
  high-order data.
\newblock {\em Journal of the American Statistical Association}, 0(0):1--34.

\bibitem[Zhang, 2010]{zhang2010}
Zhang, C.-H. (2010).
\newblock Nearly unbiased variable selection under minimax concave penalty.
\newblock {\em The Annals of statistics}, 38(2):894--942.

\bibitem[Zhou et~al., 2013a]{zhou2013tensor}
Zhou, H., Li, L., and Zhu, H. (2013a).
\newblock Tensor regression with applications in neuroimaging data analysis.
\newblock {\em Journal of the American Statistical Association},
  108(502):540--552.

\bibitem[Zhou et~al., 2013b]{zhou2013network}
Zhou, K., Zha, H., and Song, L. (2013b).
\newblock Learning social infectivity in sparse low-rank networks using
  multi-dimensional hawkes processes.
\newblock In {\em Artificial Intelligence and Statistics}, pages 641--649.

\bibitem[Zhu et~al., 2019]{zhu2019}
Zhu, X., Tang, X., and Qu, A. (2019).
\newblock Longitudinal clustering for heterogeneous binary data.
\newblock {\em Statistica Sinica}.

\bibitem[Zhu et~al., 2016]{zhu2016}
Zhu, Y., Shen, X., and Ye, C. (2016).
\newblock Personalized prediction and sparsity pursuit in latent factor models.
\newblock {\em Journal of the American Statistical Association},
  111(513):241--252.

\end{thebibliography}

\end{document}